\documentclass[a4paper,11pt]{article} 
\pdfoutput=1
\usepackage{jheppub}
\usepackage{slashed}
\usepackage[dvipsnames,table]{xcolor}
\usepackage{hyperref} %pdflatex hyperliens
\usepackage{xspace}
\usepackage[tight]{subfigure}
\usepackage{amsmath}
\usepackage{amssymb}
\usepackage{amsfonts}
\usepackage{mathrsfs}
\usepackage{comment}
\usepackage{afterpage}
\usepackage{verbatim}
\usepackage{booktabs}
\usepackage{array}
\usepackage[title,titletoc]{appendix}
\usepackage{tabularx}
\newcolumntype{C}[1]{>{\centering\arraybackslash}p{#1}}
\def\refeq#1{\mbox{(\ref{#1})}}
\def\reffi#1{\mbox{Fig.~\ref{#1}}}
\def\reffis#1{\mbox{Figs.~\ref{#1}}}
\def\refta#1{\mbox{Table~\ref{#1}}}

\def\refse#1{\mbox{Section~\ref{#1}}}
\def\refses#1{\mbox{Sections~\ref{#1}}}

\def\citere#1{\mbox{Ref.~\cite{#1}}}
\def\citeres#1{\mbox{Refs.~\cite{#1}}}

\newcommand{\newc}{\newcommand}
\newc{\beq}{\begin{equation}}
\newc{\eeq}{\end{equation}}
\newc{\bit}{\begin{itemize}}
\newc{\eit}{\end{itemize}}
\newc{\ben}{\begin{enumerate}}
\newc{\een}{\end{enumerate}}
\newc{\bce}{\begin{center}}
\newc{\ece}{\end{center}}
\newc{\bfi}{\begin{figure}}
\newc{\efi}{\end{figure}}

\newcommand\cM{{\cal M}}
\newcommand\cO{{\cal O}}

\newcommand{\ri}{\mathrm i}

\newcommand{\rT}{{\mathrm{T}}}

\newcommand{\rw}{{\mathrm{w}}}

\newcommand{\ie}{\emph{i.e.}\ }
\newcommand{\eg}{\emph{e.g.}\ }

\newcommand{\GeV}{\ensuremath{\,\text{GeV}}\xspace}
\newcommand{\TeV}{\ensuremath{\,\text{TeV}}\xspace}
\newcommand{\fb}{{\ensuremath\unskip\,\text{fb}}\xspace}

\newcommand{\PH}{\ensuremath{\text{H}}\xspace}

\newcommand{\Pp}{\ensuremath{\text{p}}}
\newcommand{\Pe}{\ensuremath{\text{e}}\xspace}
\newcommand{\Pb}{\ensuremath{\text{b}}\xspace}
\newcommand{\Pq}{\ensuremath{q}}
\newcommand{\Pt}{\ensuremath{\text{t}}\xspace}
\newcommand{\Pu}{\ensuremath{\text{u}}\xspace}
\newcommand{\Pd}{\ensuremath{\text{d}}\xspace}
\newcommand{\Ps}{\ensuremath{\text{s}}\xspace}
\newcommand{\Pc}{\ensuremath{\text{c}}\xspace}
\newcommand{\Pg}{\ensuremath{\text{g}}}

\newcommand{\PW}{\ensuremath{\text{W}}\xspace}
\newcommand{\PZ}{\ensuremath{\text{Z}}\xspace}
\newcommand{\Pl}{\ell}
                                    
\newcommand{\Mt}{\ensuremath{m_\Pt}\xspace}
\newcommand{\MH}{\ensuremath{M_\PH}\xspace}

\newcommand{\Gt}{\ensuremath{\Gamma_\Pt}\xspace}
\newcommand{\GH}{\ensuremath{\Gamma_\PH}\xspace}

\newcommand{\alphas}{\ensuremath{\alpha_\text{s}}\xspace}

\newcommand{\alphadip}{\ensuremath{\alpha_\text{dipole}}\xspace}

\newcommand{\MSbar}{\ensuremath{\overline{{\text{MS}}\xspace}}}

\newcommand{\recola}{{\sc Recola}\xspace}

\newcommand{\mocanlo}{{\sc MoCaNLO}\xspace}
\newcommand{\collier}{{\sc Collier}\xspace}

\newcolumntype{.}{D{.}{.}{-1}}
\newcolumntype{d}[1]{D{.}{.}{#1}}
\colorlet{tableoverheadcolor}{gray!37.5}
\colorlet{tableheadcolor}{gray!25}
\colorlet{tablerowcolor}{gray!12.5}

\def\draftdate{\relax}
\def\mda{\relax}
\def\mua{\relax}
\def\mla{\relax}
\def\draft{
\def\thtystars{******************************}
\def\sixtystars{\thtystars\thtystars}
\typeout{}
\typeout{\sixtystars**}
\typeout{* Draft mode!
         For final version remove \protect\draft\space in source file *}
\typeout{\sixtystars**}
\typeout{}
\def\draftdate{\today}
\def\mua{\marginpar[\boldmath\hfil$\uparrow$]%
                   {\boldmath$\uparrow$\hfil}\color{black}%
                    \typeout{marginpar: $\uparrow$}\ignorespaces}
\def\mda{\color{red}\marginpar[\boldmath\hfil$\downarrow$]%
                   {\boldmath$\downarrow$\hfil}%
                    \typeout{marginpar: $\downarrow$}\ignorespaces}
\def\mla{\marginpar[\boldmath\hfil$\rightarrow$]%
                   {\boldmath$\leftarrow $\hfil}%
                    \typeout{marginpar: $\leftrightarrow$}\ignorespaces}
\def\Mua{\marginpar[\boldmath\hfil$\Uparrow$]%
                   {\boldmath$\Uparrow$\hfil}\color{black}%
                    \typeout{marginpar: $\uparrow$}\ignorespaces}
\def\Mda{\color{red}\marginpar[\boldmath\hfil$\Downarrow$]%
                   {\boldmath$\Downarrow$\hfil}%
                    \typeout{marginpar: $\downarrow$}\ignorespaces}
\def\Mla{\marginpar[\boldmath\hfil\textcolor{red}{$\Rightarrow$}]%
                   {\boldmath\textcolor{red}{$\Leftarrow $}\hfil}%
                    \typeout{marginpar: $\leftrightarrow$}\ignorespaces}
\overfullrule 5pt
\oddsidemargin 15mm
\marginparwidth 29mm
}

\newcommand{\mc}{\mathcal}
\let\nnb\notag

\let\Mw\MW

\let\Mz\MZ

\let\Mwo\MWOS
\let\Gwo\GWOS
\let\Mzo\MZOS
\let\Gzo\GZOS
\let\ci\ri
\newcommand{\Mvo}{M^{\rm OS}_{V}}
\newcommand{\Mv} {M_{V}}
\newcommand{\Gvo}{\Gamma^{\rm OS}_{V}}
\newcommand{\Gv} {\Gamma_{V}}
\let\as\alphas
\newcommand{\pt}[1]{p_{\rT,{#1}}}
\newcommand{\tm}[1]{M_{\rT,{#1}}}

\title{NLO QCD corrections to off-shell $\text{t}\overline{\text{t}}\text{W}^+$ production at the LHC}
\author{Ansgar Denner and}
\author{Giovanni Pelliccioli}
\affiliation{Universit\"at W\"urzburg, Institut f\"ur Theoretische Physik und Astrophysik, 97074 W\"urzburg, Germany}
\emailAdd{ansgar.denner@physik.uni-wuerzburg.de}
\emailAdd{giovanni.pelliccioli@physik.uni-wuerzburg.de}
\date{\draftdate}

\abstract{We present results of a computation of NLO QCD corrections
  to the production of an off-shell top--antitop pair in association
  with an off-shell $\PW^+$ boson in proton--proton collisions. As the
  calculation is based on the full matrix elements for the process
  $\Pp\Pp\to
  {\Pe}^+\nu_{\Pe}\,\mu^-\overline{\nu}_\mu\,\tau^+\nu_\tau\,{\Pb}\,\overline{\Pb}$,
  all off-shell, spin-correlation, and interference effects are
  included. The NLO QCD corrections are about $20\%$ for the
  integrated cross-section. Using a dynamical scale, the corrections to
  most distributions are at the same level, while some distributions
  show much larger $K$-factors in suppressed regions of phase space.
  We have performed a second calculation based on a double-pole
  approximation. While the corresponding results agree with the full
  calculation within few per cent for integrated cross-sections, the
  discrepancy can reach $10\%$ and more in regions of phase space that
  are not dominated by top--antitop production.  As a consequence,
  on-shell calculations should only be trusted to this level of
  accuracy.  }


\begin{document}

\maketitle
%%%%%%%%%%%%%%%%%%%%%
%\tableofcontents
\section{Introduction}\label{sec:intro}
The production of top--antitop-quark pairs in association with a
massive electroweak vector boson is of great importance at the Large
Hadron Collider (LHC), both as a test of the Standard Model (SM) and
for the search for effects of new physics.  The presence of a dominant
resonance structure that features three massive unstable particles
makes such a process one of the heaviest signatures that can be
detected and investigated at the LHC.

Both $\Pt \overline{\Pt}\PW^\pm $ and $\Pt\overline{\Pt}\PZ$
production have been observed and studied by the ATLAS and CMS
experiments with $8\TeV$
\cite{Aad:2015eua,Khachatryan:2015sha} and $13\TeV$ centre-of-mass (CM)
energy data
\cite{Aaboud:2016xve,Sirunyan:2017uzs,Aaboud:2019njj,CMS:2019too}. In
particular, the ATLAS collaboration has measured
${\Pt\overline{\Pt}}\PW^\pm$ production in the three-charged-leptons
channel \cite{Aaboud:2019njj}, which is the process considered in this
paper.  The collected data of the LHC Runs 1 and 2 and the 
luminosities expected in the coming runs will deliver an improved
understanding of the dynamics of $\Pt\overline{\Pt}V$ processes.

The study of $\Pt\overline{\Pt}V$ production is expected to provide
a direct access to the top-quark coupling to gauge bosons, and
specifically to possible deviations from the SM prediction due
to new physics \cite{Dror:2015nkp,Buckley:2015lku,Bylund:2016phk}.
The $\Pt\overline{\Pt}\PW^\pm$ and $\Pt\overline{\Pt}\PZ$ signatures
can be directly affected by beyond-Standard Model (BSM) effects, for
example by the presence of supersymmetry
\cite{Guchait:1994zk,Barnett:1993ea}, vector-like quarks
\cite{Aguilar-Saavedra:2013qpa}, little Higgs bosons
\cite{Perelstein:2005ka} and technicolour \cite{Chivukula:1994mn}.
%%%%%%
The SM $\Pt\overline{\Pt}\PZ$ production, with $\PZ\rightarrow
\nu\overline{\nu}$, represents an important background for dark-matter
signals \cite{Bevilacqua:2019cvp}.

In particular, the associated production of top quarks with a $\PW^\pm$
boson in the fully-leptonic decay channel represents a rare process at
the LHC, owing to the presence of two like-sign leptons in the final
state. Therefore, despite a relatively low rate, it constitutes an
optimal signature for BSM searches, with a particular focus on
supersymmetry \cite{Barnett:1993ea,Guchait:1994zk}, supergravity
\cite{Baer:1995va}, Majorana neutrinos \cite{Almeida:1997em} and
modified Higgs sectors \cite{Maalampi:2002vx,Contino:2008hi}.
The $\Pt\overline{\Pt}\PW^\pm$ signature could noticeably improve the
investigation of the $\Pt\overline{\Pt}$ charge asymmetry
\cite{Maltoni:2014zpa}, that is expected to be larger than in pure
$\Pt \overline{\Pt}$ production, due to the absence of neutral initial
states at leading order (LO) and next-to-leading order (NLO).
Moreover, the study of polarisation-related observables in
$\Pt\overline{\Pt}\PW^\pm$ production provides additional sensitivity to BSM
effects \cite{Maltoni:2014zpa}.

The $\Pt\overline{\Pt}V$ production is also an important background
for top--antitop production in association with a Higgs boson
\cite{Maltoni:2015ena}, which has been recently observed
\cite{Aaboud:2018urx,Sirunyan:2018hoz} and further investigated in the
multi-lepton channel \cite{ATLAS-CONF-2019-045,CMS-PAS-HIG-17-004} at
the LHC.

Both in the direct measurement of $\Pt\overline{\Pt}\PW^\pm$
production \cite{Sirunyan:2017uzs,Aaboud:2019njj} and in the
measurement of the $\Pt\overline{\Pt}\PH$ signal
\cite{ATLAS-CONF-2019-045,CMS-PAS-HIG-17-004}, a tension between
$13\TeV$ CM data and the SM predictions has been observed in the
$\Pt\overline{\Pt}\PW^\pm$ modelling, both in the inclusive
cross-section, and in kinematic regimes where
$\Pt\overline{\Pt}\PW^\pm$ production is expected to be the leading
contribution.  In order to address such a tension with the data and to
allow for more precise investigations, improved precision and accuracy
are needed for the theoretical description of this process.

A number of phenomenological studies are already present in the
literature. The hadronic production of $\Pt\overline{\Pt}\PW^\pm$ was
first computed with NLO QCD accuracy for LHC energies including
top-quark decays with full spin correlations, using decay chains to
simulate the semileptonic final state \cite{Campbell:2012dh}.
Parton-shower effects of both $\Pt\overline{\Pt}\PW^\pm$ and
$\Pt\overline{\Pt}\PZ$ production with semileptonic top decays were
added on top of NLO QCD predictions for the two- ($\PW$) and
three-charged-leptons ($\PZ$) final states at the LHC at $7/8\TeV$ CM
energy \cite{Garzelli:2012bn}.  The importance of this processes for the
study of the $\Pt\overline{\Pt}$ charge asymmetry
\cite{Maltoni:2014zpa} and its impact on $\Pt\overline{\Pt}\PH$
searches \cite{Maltoni:2015ena} was examined at NLO QCD in inclusive
production. The impact of electroweak corrections on
$\Pt\overline{\Pt}\PW^\pm$ production was investigated in inclusive
production (no decays of the three resonances) in
\citeres{Frixione:2015zaa,Frederix:2017wme}.  Resummed calculations
for the on-shell production of $\Pt\overline{\Pt}$ in association with
a heavy boson were performed up to next-to-next-to-leading-logarithmic
(NNLL) accuracy
\cite{Li:2014ula,Broggio:2016zgg,Kulesza:2018tqz,Broggio:2019ewu,Kulesza:2020nfh}.
An NLO QCD analysis including parton-shower matching with the focus of
effects of spin correlations and subleading electroweak corrections on
asymmetries was presented in \citere{Frederix:2020jzp}.
The first computation of NLO QCD corrections including
all off-shell effects and spin correlations appeared  very recently
 \cite{Bevilacqua:2020pzy} for hadronic collisions at $13\TeV$ CM energy
in the three-charged-leptons decay channel.

The goal of this paper is to present an independent computation of the
complete NLO QCD corrections to the off-shell production of
$\Pt\overline{\Pt}\PW^+$ in the three-charged-leptons decay channel.
The specific final state we consider, the scale choices we analyse,
and the SM input parameters differ from those of
\citere{Bevilacqua:2020pzy}.
However, we have reproduced the results of \citere{Bevilacqua:2020pzy}
using the same setup employed therein, finding very good agreement both at the
integrated and at the differential level.

This work is organised as follows.  In \refse{sec:calcdetails} we
present details of our calculation of the
${\Pt\overline{\Pt}\PW^+}$ process, including a description of real
and virtual NLO corrections. We describe an approximated calculation
of the same process that relies on double-pole-approximation
techniques and a number of numerical checks we have performed to
validate the full off-shell computation.  In \refse{sec:numresults} we
present the results of numerical simulations both at the integrated
and at the differential level.  After introducing the setup for
the numerical simulations in \refse{sec:input}, we discuss the scale
choice in \refses{sec:integrated} and \ref{subsec:scalediff}. 
In \refse{subsec:ourscale} we analyse a number of kinematic variables,
including both directly measurable observables at the LHC and
kinematic variables based on Monte Carlo truth.  The results obtained from
full matrix elements are compared with those in the double-pole
approximation in \refse{se:fulvsdpa}.  We draw our conclusions in
\refse{sec:conclusions}.

\section{Details of the calculation}\label{sec:calcdetails}

In this paper we present the NLO QCD corrections to the process
\beq\label{eq:procdef}
\Pp\Pp\to {\Pe}^+\nu_{\Pe}\,\mu^-\overline{\nu}_\mu\,\tau^+\nu_\tau\,{\Pb}\,\overline{\Pb}\,.
\eeq
The calculation has been performed with \mocanlo, a Monte Carlo
generator that has been used to simulate other processes involving top
quarks with NLO EW and QCD accuracy
\cite{Denner:2015yca,Denner:2016jyo,Denner:2017kzu,Denner:2016wet}.
It is interfaced with \recola \cite{Actis:2012qn, Actis:2016mpe} which
provides the tree-level and one-loop SM amplitudes using the \collier
library \cite{Denner:2016kdg} to perform the reduction and numerical
evaluation of one-loop integrals
\cite{Denner:2002ii,Denner:2005nn,Denner:2010tr}.

The leading tree-level contributions to the considered process are of
order $\cO (\as^2\alpha^6)$, and come uniquely from quark-induced
partonic channels. Since we assume a diagonal quark-mixing matrix, the
only possible initial states are ${\Pu}\bar{\Pd}$ and
${\Pc}\bar{\Ps}$.  Similarly to $\Pt \overline{\Pt}$ production, other
LO contributions are either suppressed [$\cO (\alpha^8)$] or
vanishing thanks to colour algebra [$\cO (\as\alpha^7)$] and are
neglected here.  No photon- or gluon-initiated processes contribute to
this process at LO.

In the following we focus on the NLO QCD corrections to the leading
tree-level contribution, leaving NLO EW corrections to future work.
All resonant and non-resonant diagrams of order $\cO
(g_{\mathrm{s}}^2\,g^6)$ (Born, tree-level), $\cO
(g_{\mathrm{s}}^3\,g^6)$ (real, tree-level) and $\cO
(g_{\mathrm{s}}^4\,g^6)$ (virtual, one-loop) are included in the
calculation, in order to account for all interferences and off-shell
effects related to the electroweak gauge bosons and top quarks.

\begin{figure}[t]
  \centering
  \subfigure[Two top, three W\label{subfig:2t}]{\includegraphics[scale=0.35]{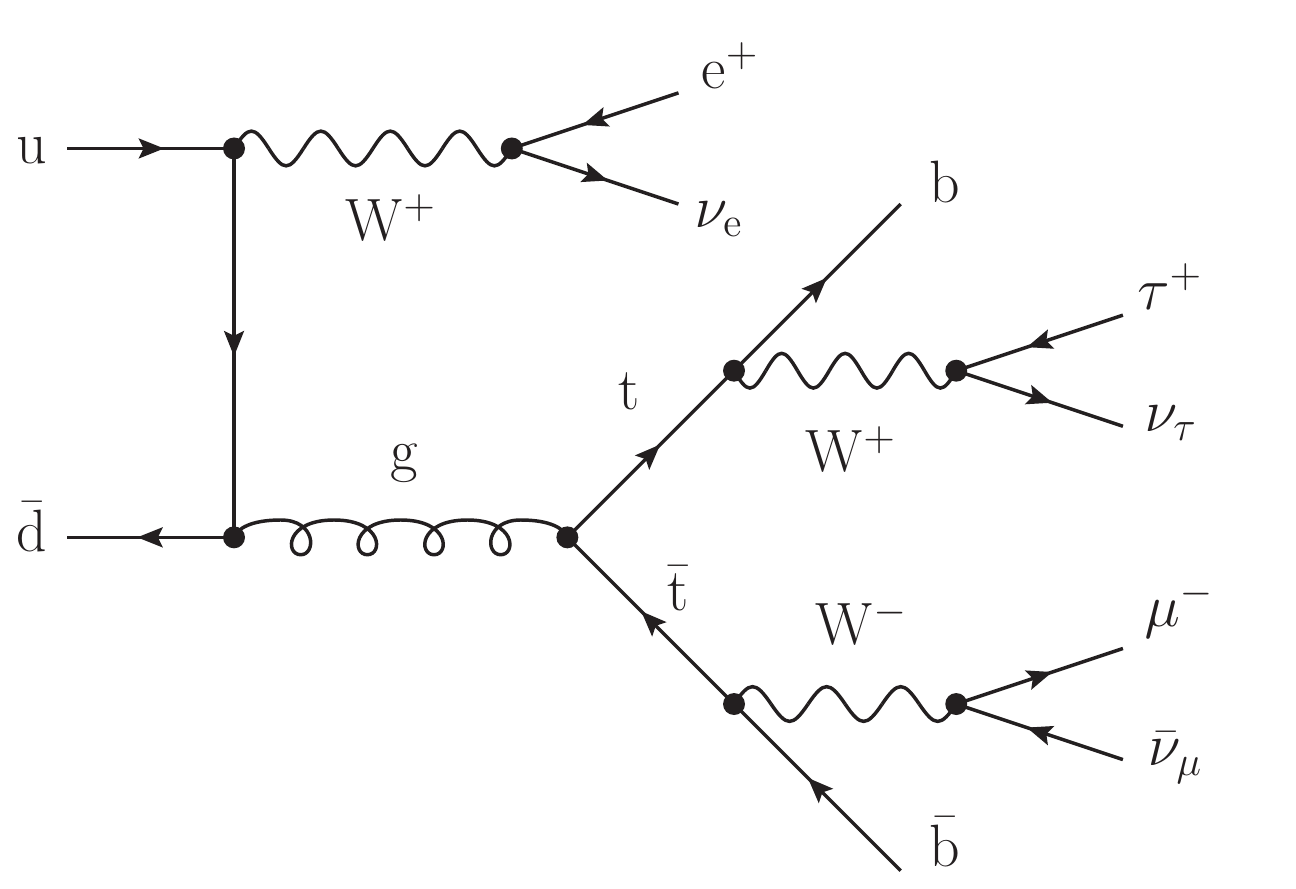}}
  \subfigure[One top, three W\label{subfig:1t}]{\includegraphics[scale=0.35]{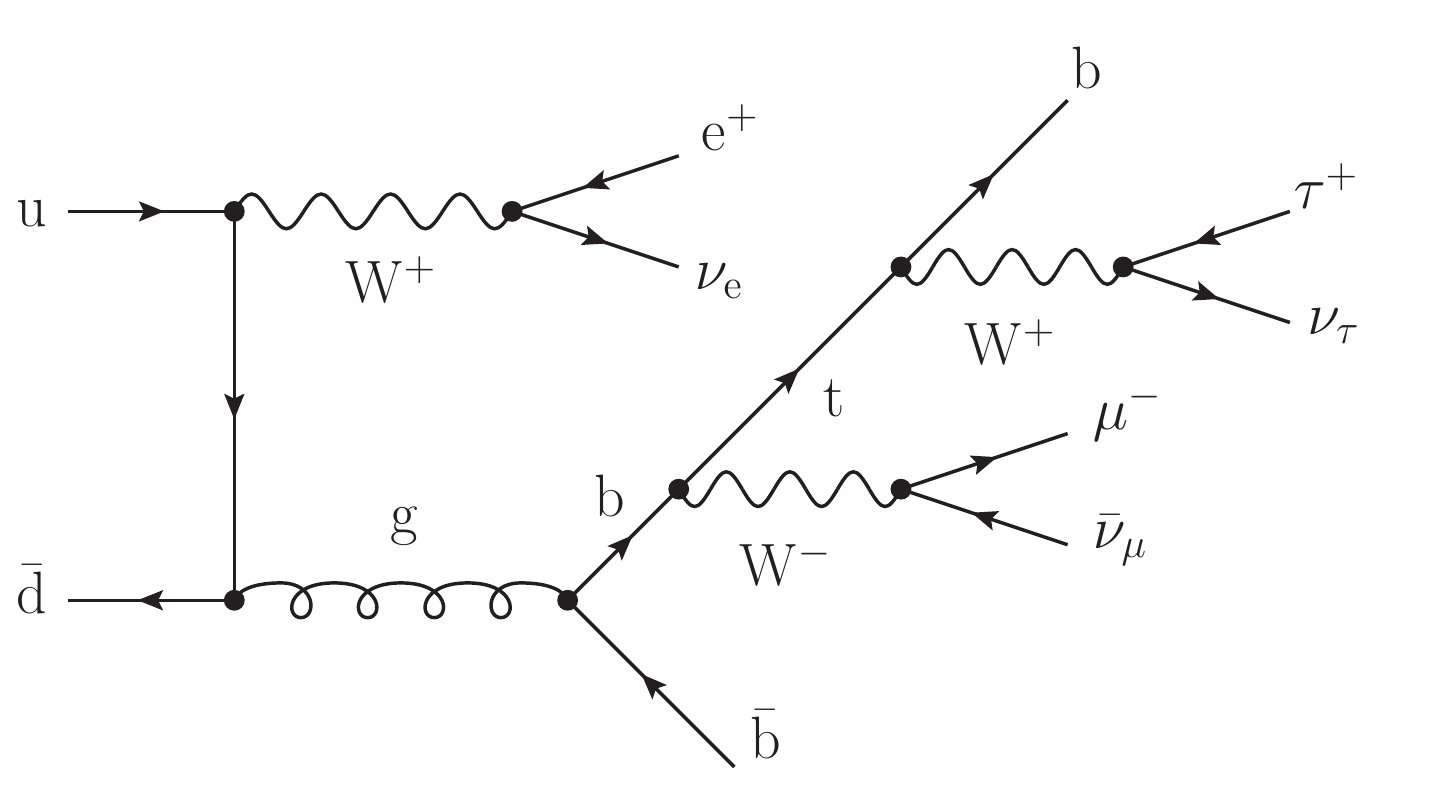}}  
  \subfigure[Zero top, three W, Higgs\label{subfig:0t3w1h}]{\includegraphics[scale=0.35]{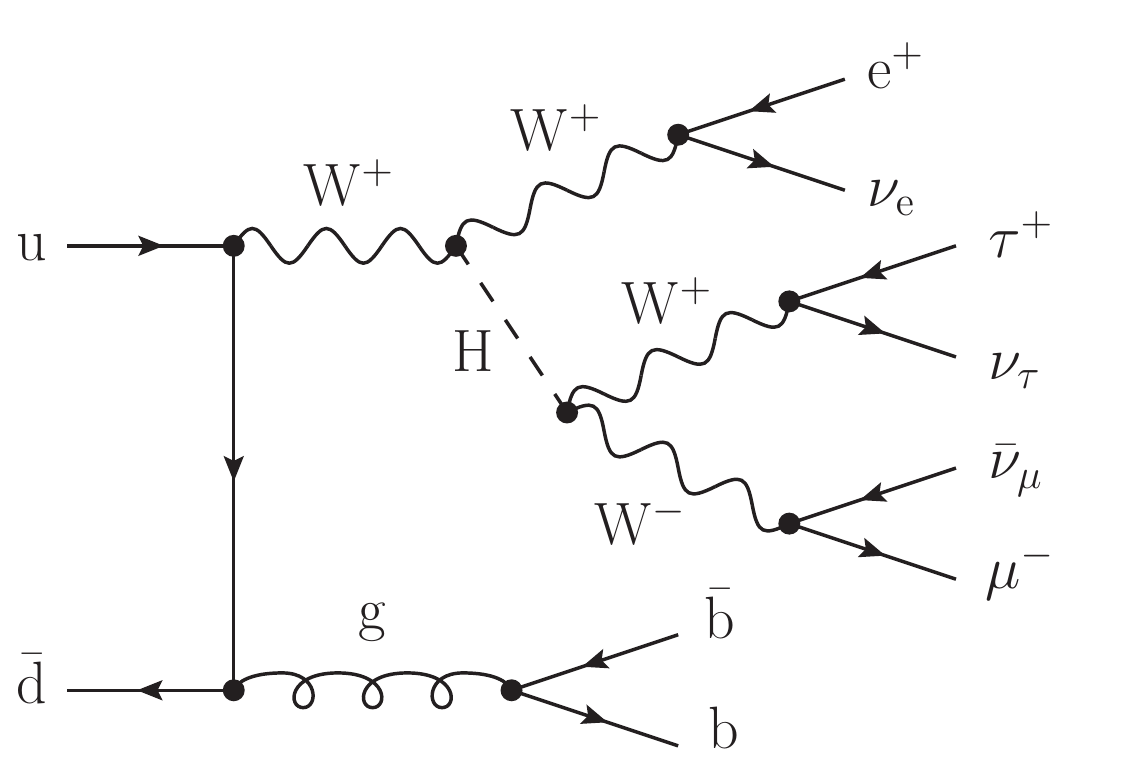}}
  \subfigure[Zero top, three W, \PZ\label{subfig:0t3w}]{\includegraphics[scale=0.35]{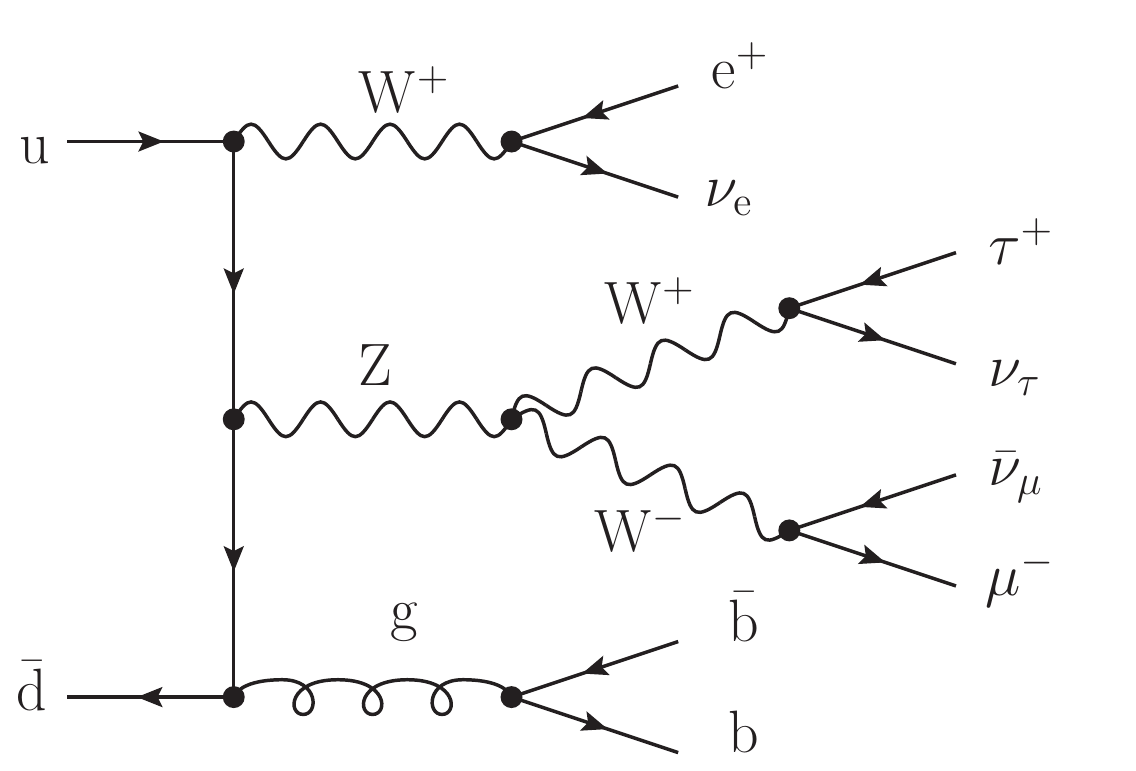}}
  \subfigure[Zero top, two W\label{subfig:0t2w}]{\includegraphics[scale=0.35]{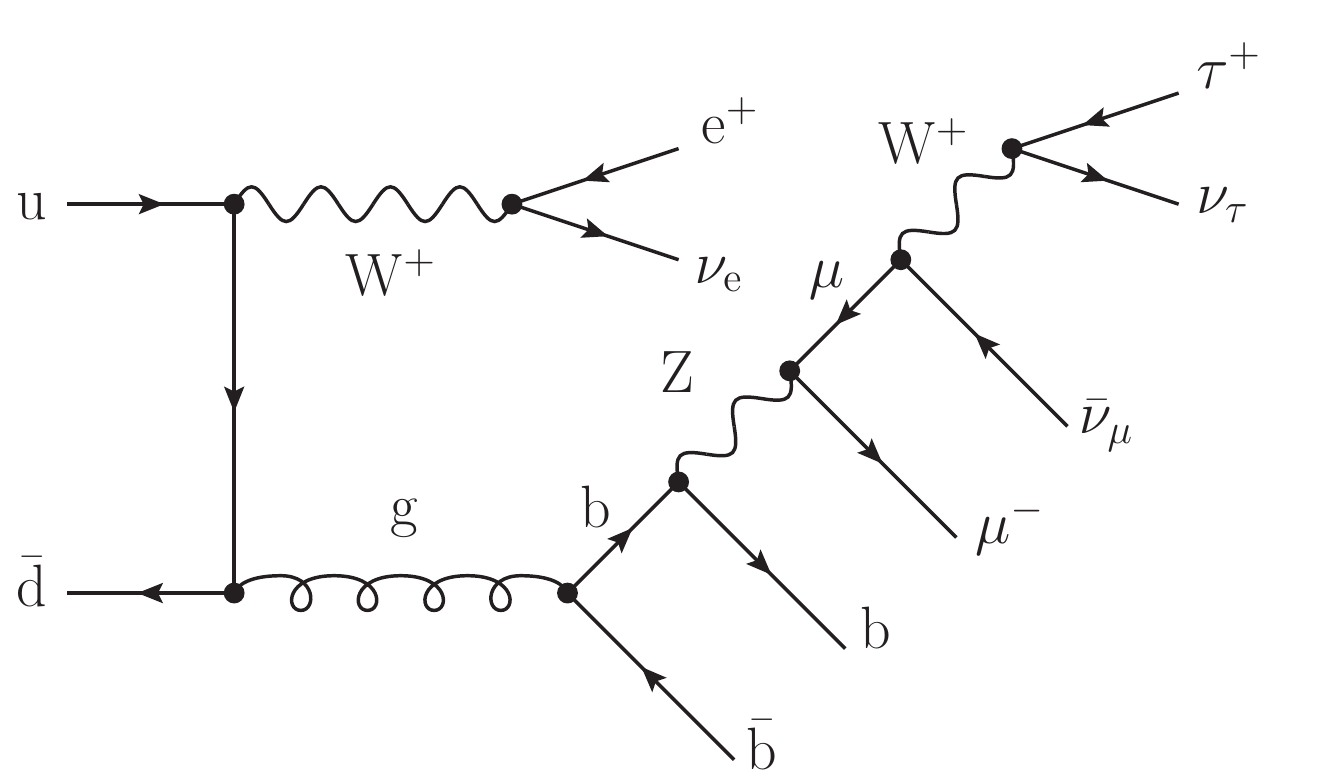}}
  \subfigure[Zero top, one W\label{subfig:0t1w}]{\includegraphics[scale=0.35]{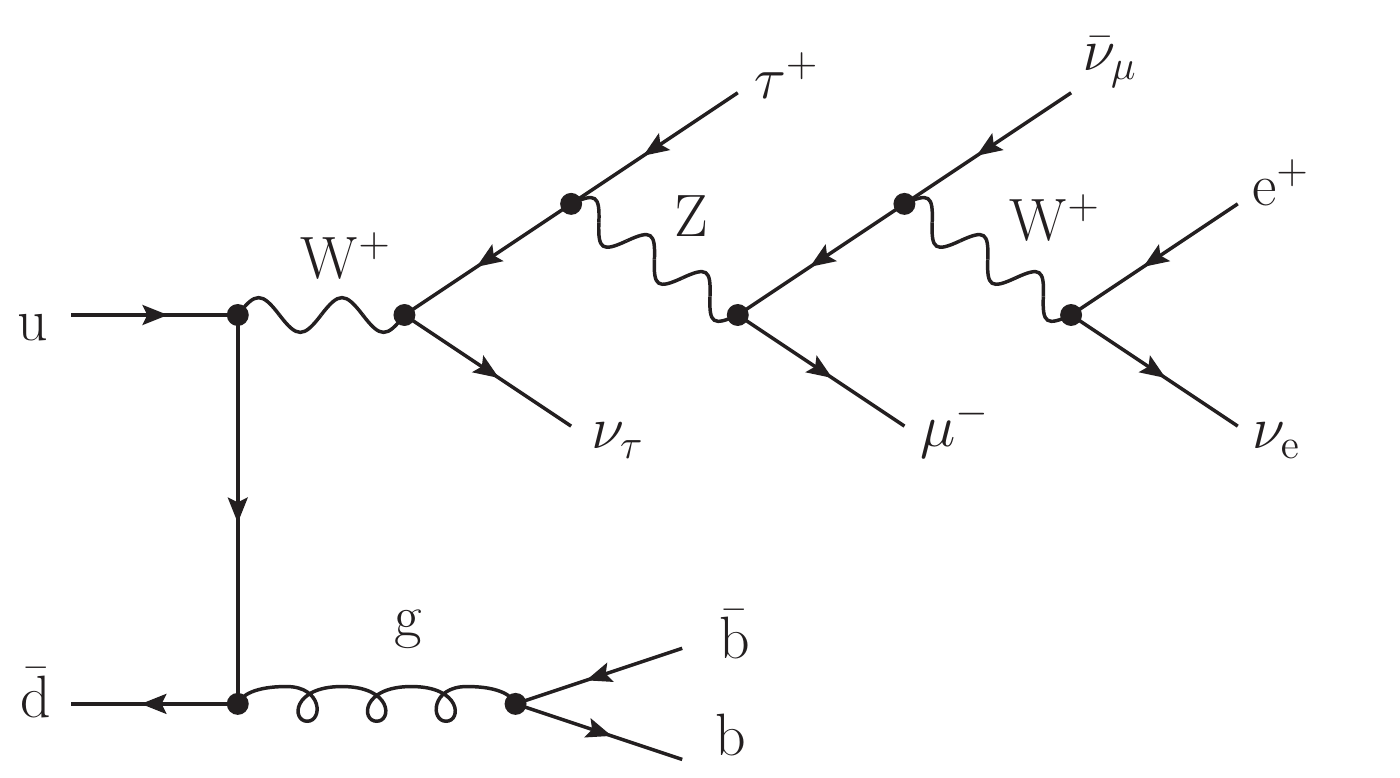}}
  \caption{Sample tree-level diagrams contributing to ${\Pt\overline{\Pt}}\PW^+$ production at the LHC.}\label{born_diags}
\end{figure}

At Born level, 286 diagrams contribute to each partonic process.
The large number of final-state particles allows for several resonance
structures, In \reffi{born_diags} we show a few sample diagrams that
contribute at LO and feature two, one, or zero resonant top/antitop
quarks.
In most of the diagrams, the propagating gluon is radiated from
initial-state quarks and converts to ${\Pb\overline{\Pb}}$ [see for
example \reffis{subfig:0t3w1h}, \ref{subfig:0t3w} and
\ref{subfig:0t1w}].  However such diagrams give a small contribution
to the squared amplitude, since they do not contain resonant top or
antitop quarks. The leading contributions are provided by diagrams
with a resonant ${\Pt\overline{\Pt}}$ pair, like the one depicted in
\reffi{subfig:2t}. Also single-top-resonant diagrams like the one in
\reffi{subfig:1t} yield a visible contribution at LO.  Other diagram
topologies give minor contributions: a ${\PW^+\PW^-}$ pair can be
produced via the decay of a Higgs [\reffi{subfig:0t3w1h}] or a
$\PZ$~boson [\reffi{subfig:0t3w}]. We observe that all diagrams
contain at least one resonant $\PW^+$ boson.

\subsection{Real corrections}
Given the LO structure of the process, six possible partonic channels
contribute to the real corrections at order $\cO(\as^3\alpha^6)$,
\begin{align}\label{realchannels}
\Pu\bar{\Pd}/\Pc\bar{\Ps} \to{} & \Pe^+\nu_\Pe\,\mu^-\bar{\nu}_\mu\,\tau^+\nu_\tau\,\Pb\,\bar\Pb\,\Pg\,\nnb\\
\Pu\Pg/\Pc\Pg \to {}& \Pe^+\nu_\Pe\,\mu^-\bar{\nu}_\mu\,\tau^+\nu_\tau\,\Pb\,\bar\Pb\,\Pd/\Ps\,\nnb  \\
\Pg\bar{\Pd}/\Pg\bar{\Ps} \to{} & \Pe^+\nu_\Pe\,\mu^-\bar{\nu}_\mu\,\tau^+\nu_\tau\,\Pb\,\bar\Pb\,{\bar{\Pu}/\bar{\Pc}}\,.
\end{align}
Since we consider a final state with three charged leptons, the
additional gluon can only be radiated off the initial-state quark
lines, the final-state $\Pb$~jets, or internal light-quark, bottom and
top-quark propagators. 
For each real-radiation channel 1916 Feynman diagrams contribute. 
The $\Pq\bar{\Pq}'$ channel is the one featuring
the largest number of infrared (IR) singular regions, and thus is the most 
demanding in the numerical calculation.  We treat bottom quarks as
massless particles and work in the five-flavour scheme. For the final
states of the processes \refeq{realchannels}, no initial states
involving bottom quarks contribute for a diagonal quark-mixing matrix
that we assume.

We treat the IR singularities with the Catani--Seymour subtraction
scheme \cite{Catani:1996vz}. The initial-state collinear QCD
singularities are absorbed in the PDFs via the $\MSbar$ scheme. The
spin-correlated and colour-correlated matrix elements entering the
unintegrated subtraction counterterms are computed with \recola
\cite{Actis:2012qn, Actis:2016mpe}.  As a validation of the correct
computation of the subtracted real and integrated counterterms, we
have varied the $\alphadip$ parameters \cite{Nagy:2003tz} in the
definition of the dipoles, as is further explained in
\refse{se:valid}.

Although computing the virtual contributions involves the evaluation
of complicated one-loop diagrams (see the following section), the
numerical simulation of the (subtracted) real contributions to NLO
observables represents the most CPU-intensive part of the calculation,
in spite of its tree-level nature and the relatively small number of
QCD subtraction dipoles.  This is due to the large multiplicity of the
final state (three coloured partons, six leptons) but also due to the
relatively small numerical contribution of the virtual corrections.
Tackling the NLO EW corrections for the same process will be even more
challenging, as the number of QED dipoles is larger than the one of
QCD dipoles.

\subsection{Virtual corrections}\label{subsec:virt}

\begin{figure}[t]
  \centering
  \includegraphics[scale=0.35]{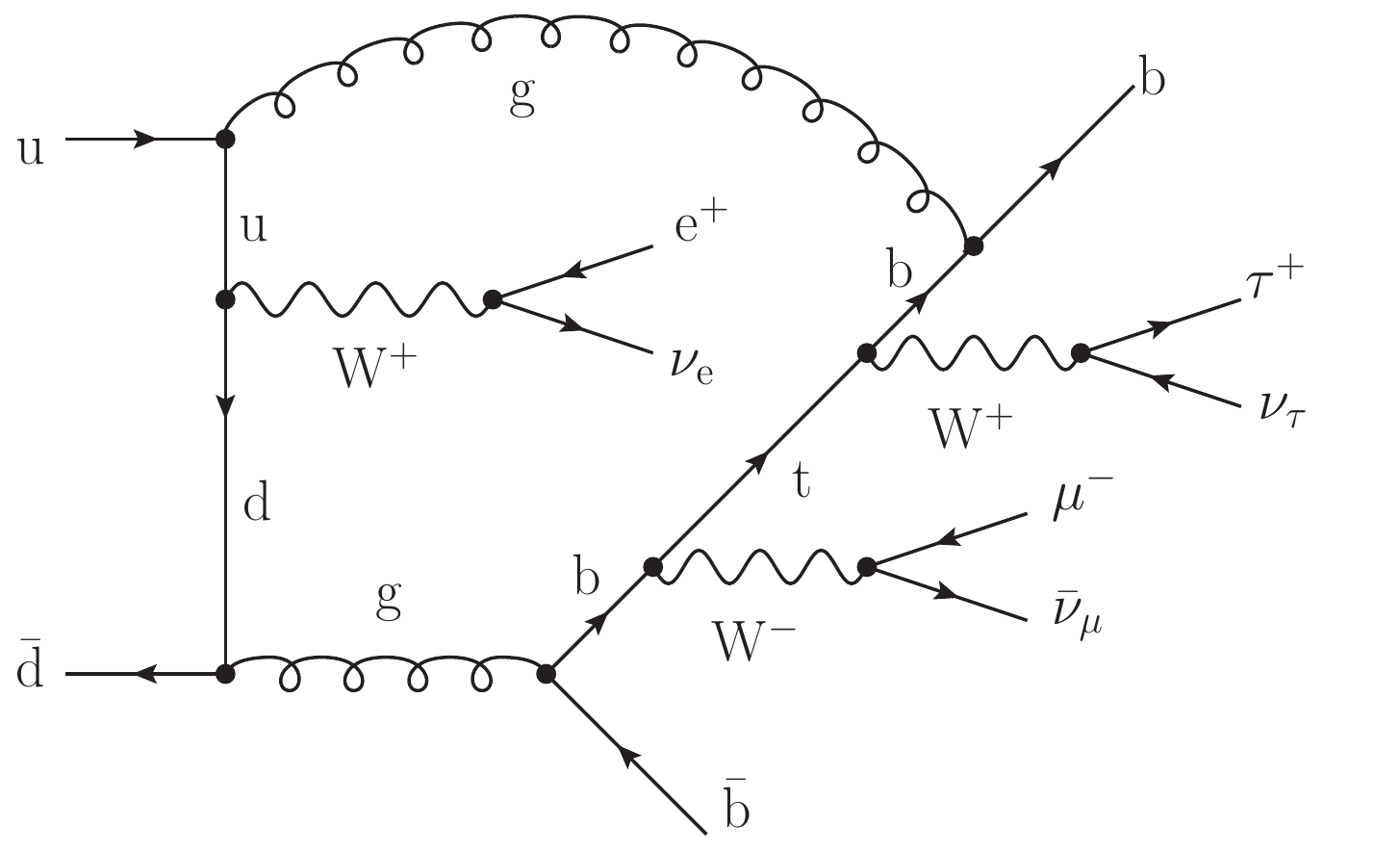} 
  \caption{Sample one-loop diagram contributing to hadronic
    ${\Pt\overline{\Pt}}\PW^+$ production at NLO QCD 
    involving a 7-point function.}\label{virtual7point}
\end{figure}

The virtual QCD corrections are obtained by inserting 
a gluon in the tree-level amplitudes, resulting in roughly
6000 Feynman diagrams for each partonic channel.  
The most complicated one-loop
contributions involve 7-point functions appearing when a virtual
gluon connects the initial-state light quarks and a final-state
b~quark in diagrams which do not feature the $\Pt\overline{\Pt}$
resonance structure. An example is depicted in \reffi{virtual7point}.
The virtual corrections to the considered process are computed with SM
amplitudes provided by \recola \cite{Actis:2012qn, Actis:2016mpe}
interfaced to the \collier library \cite{Denner:2016kdg} for the
numerical evaluation of scalar and tensor one-loop integrals. All 
unstable particles are treated in the complex-mass scheme
\cite{Denner:1999gp,Denner:2000bj,Denner:2005fg,Denner:2006ic},
resulting in complex input values for the electroweak boson masses,
the top-quark mass, and the electroweak mixing angle,
\beq
%%%%
\mu_V^2 = \Mv^2-\ri\Gv \Mv\quad (V={\PW,\PZ})\,,\qquad 
\mu_\Pt^2 = \Mt^2-\ri\Gt \Mt\,,\qquad 
\cos^2\theta_{\rw} = \frac{\mu_{\PW}^2}{\mu_{\PZ}^2}\,.
%%%%
\eeq

\subsection{Double-pole approximation}\label{dpa}
The hadronic process defined in Eq.~\refeq{eq:procdef} is dominated by
the resonant production of a top--antitop-quark pair in association
with a $\PW^+$ boson. Most of the results that are present in the
literature \cite{Frederix:2017wme, Frixione:2015zaa, Maltoni:2014zpa,
  Maltoni:2015ena, Broggio:2016zgg, Kulesza:2018tqz, Kulesza:2020nfh,
  Broggio:2019ewu} are inclusive, \ie they rely on the
on-shell production of the three massive particles.  More refined
techniques have been investigated to complement the inclusive results
with spin correlations and off-shell effects, without computing the
full off-shell process, \eg using the narrow-width approximation
 \cite{Campbell:2012dh,Frederix:2020jzp}.  Such approximations help
in assessing the importance of resonance structures as well as in
validating parts of the full calculation.

A universal technique for processes dominated by a pair of resonances
is the so-called double-pole approximation (DPA), which has been
widely used in the literature for the calculation of NLO EW virtual
corrections to several processes, like di-boson
\cite{Denner:1997ia,Jadach:1996hi,
  Jadach:1998tz,Beenakker:1998gr,Billoni:2013aba,Denner:2000bj},
vector-boson scattering \cite{Biedermann:2016yds} and top--antitop
pair production \cite{Denner:2016jyo}.  In general, the
pole approximation \cite{Stuart:1991xk,Denner:2019vbn} amounts to
selecting all diagrams containing the resonance structure that is
expected to be dominant in the full process. The Breit--Wigner
modulation of the selected resonances is preserved with off-shell
kinematics, while in the rest of the amplitude the resonant particles
are treated with on-shell kinematics, obtained by means of properly
defined projections. While the DPA is applied to the virtual
contributions, the Born and real contributions to the NLO
calculation are usually computed with full matrix elements
\cite{Denner:2000bj}. The extension of this technique to QCD
corrections is straightforward.

\begin{figure}
  \centering
  \hspace{0.3cm}\includegraphics[scale=0.35]{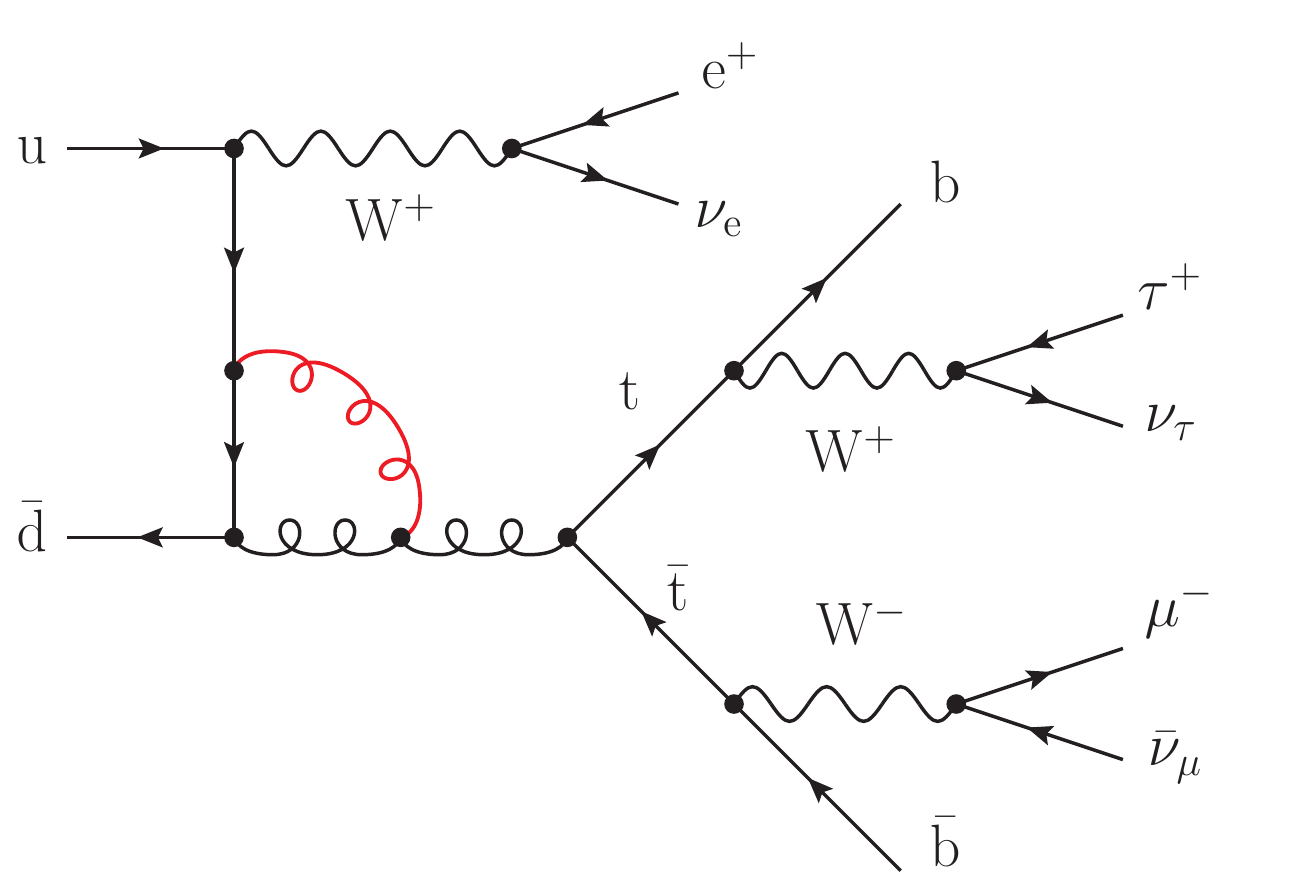}
  \includegraphics[scale=0.35]{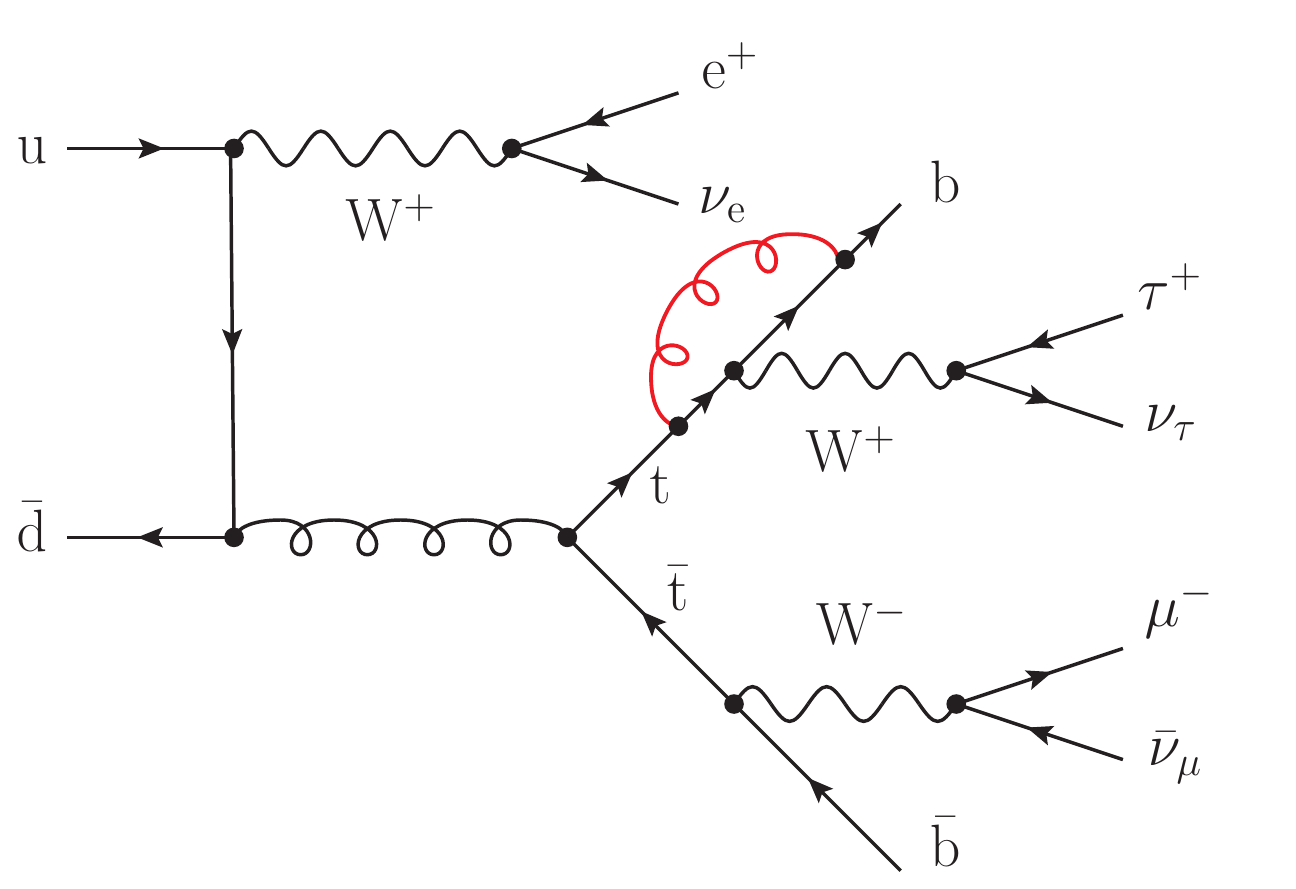}
  \includegraphics[scale=0.35]{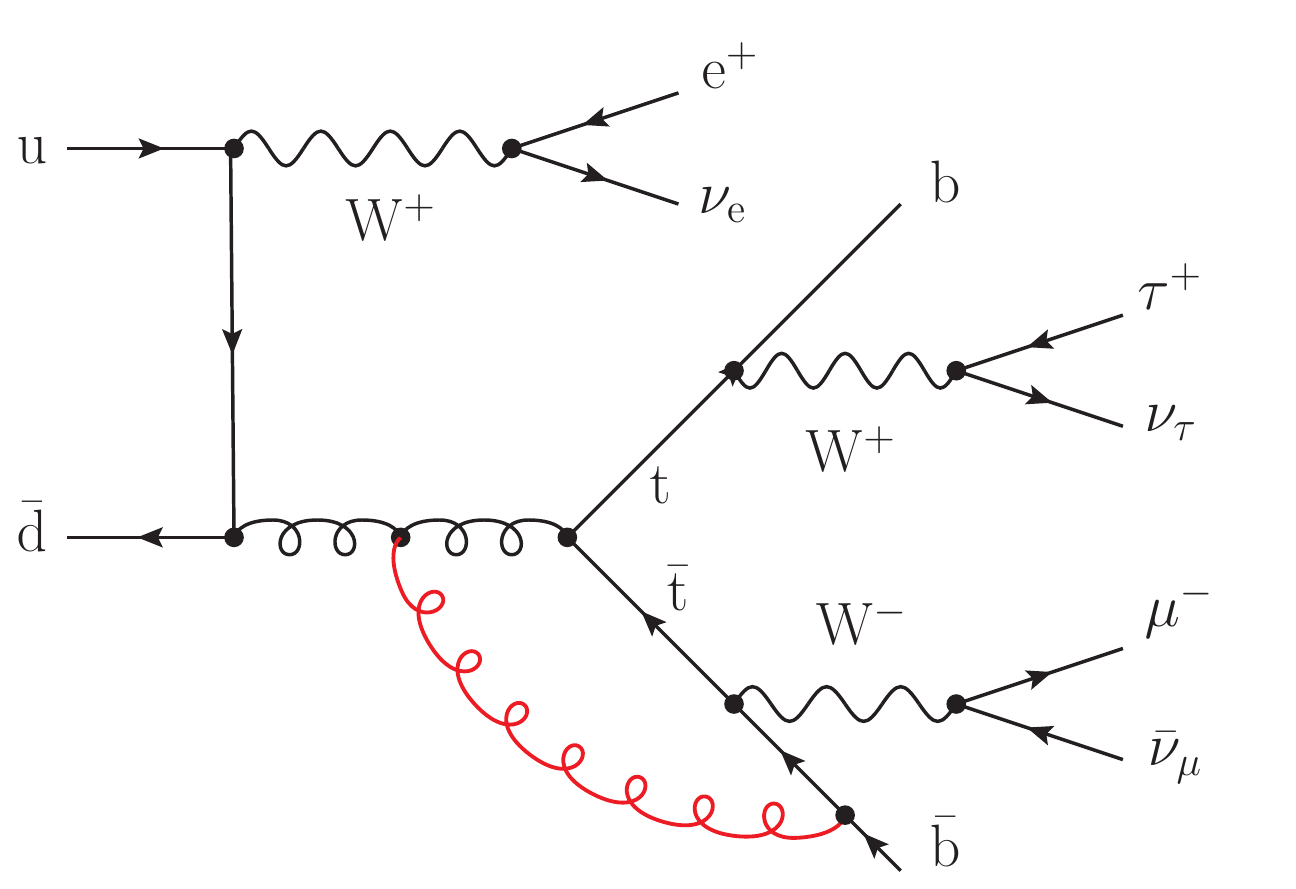}
  \caption{Sample diagrams for virtual corrections to hadronic
    ${\Pt\overline{\Pt}}\PW^+$ production: 
    factorisable corrections to the top-quark production (left)
    and decay (middle), non-factorisable corrections
    (right).}\label{factnonfact}
\end{figure}

We have applied a DPA to the top and antitop resonances in order to
mimic as much as possible the on-shell calculations that are available
in the literature for ${\Pt\overline{\Pt}\PW}$ production.  The QCD
corrections affect top quarks, being coloured particles, both in the
production and in the decay parts of the amplitude.  Therefore, both
factorisable and non-factorisable gluonic corrections contribute to
the resonant production of a ${\Pt\overline{\Pt}}$ pair.  The
factorisable corrections can be written in the general form
\begin{multline}\label{eq:fact}
          \cM^{I\rightarrow F_a F_b}_{\rm virt, fact}= \\
\sum_{h_a,h_b}         
\frac{\cM^{I\rightarrow a\,b}_{\rm virt}\,\cM^{a\rightarrow F_a}_{\rm LO}\,\cM^{b\rightarrow F_b}_{\rm LO}+
  \cM^{I\rightarrow a\,b}_{\rm LO}\,\cM^{a\rightarrow F_a}_{\rm virt}\,\cM^{b\rightarrow F_b}_{\rm LO} +
  \cM^{I\rightarrow a\,b}_{\rm LO}\,\cM^{a\rightarrow F_a}_{\rm LO}\,\cM^{b\rightarrow F_b}_{\rm virt}
}{\left(s_{a}-\Mt^2+\ci \Mt\Gt\right)\,\left(s_{b}-\Mt^2+\ci \Mt\Gt\right)}\,, 
\end{multline}
where $s_a,s_b$ and $h_a,h_b$ stand for the virtualities and helicities of
the intermediate top and antitop quarks, respectively.  Further,
$\cM^{I\rightarrow a\,b}$, $\cM^{a\rightarrow F_a}$, $\cM^{b\rightarrow
  F_b}$ are the $\Pt\overline{\Pt}$ production amplitude and the top-
and antitop-quark decay amplitudes.  The non-factorisable corrections
to the resonant amplitude have a universal structure
\cite{Denner:1997ia,Accomando:2004de,Dittmaier:2015bfe},
\beq\label{eq:nfact}
2 \,{\rm Re}\left(\cM_{\rm LO}^*\,\cM_{\rm virt,nfact}\right)
=  \left|\cM_{\rm LO}\right|^2\,\delta_{\rm nfact}\,,
\eeq
and are needed for the complete subtraction of IR singularities, since
the real contributions are not treated with the pole approximation
\cite{Denner:1997ia}.  Sample diagrams for factorisable and
non-factorisable QCD corrections are shown in \reffi{factnonfact}.
The on-shell kinematics for the resonant contributions is obtained by
suitable projections of the momenta, which involve some ambiguity.
We employ on-shell projections for top quarks that preserve the
off-shell-ness of the corresponding $\PW$ bosons, in the same fashion
as in \citere{Denner:2014wka}.

In $\Pt\overline{\Pt}\PW^+$ production two virtual $\PW^+$ bosons
appear, one of which results from the top-quark decay. Thus, the DPA
receives two distinct contributions, one where the top quark decays
into $\Pe^+\nu_\Pe\Pb$ and one where it decays into $\tau^+\nu_\tau\Pb$.
Since the interference of these contributions is not doubly resonant,
the full contributions in DPA are simply obtained upon adding these
two contributions.

One last remark concerns the definition of the part of the NLO
corrections that is calculated in the DPA.  In the original DPA
computations \cite{Denner:2000bj} and in most computations performed
with \mocanlo \cite{Denner:2016jyo,Denner:2016wet,Biedermann:2016yds}
the DPA has been applied to the finite sum of the virtual corrections and the
$I$-operator of the integrated Catani--Seymour dipoles, while the $P$-
and $K$-operators have been calculated with off-shell kinematics. 
In this way, the infrared singularities are cancelled exactly,
but the finite parts become scheme dependent. Nonetheless, the scheme
dependence is of the order of the intrinsic uncertainty of the DPA 
(see Section 2.3 of \citere{Denner:2000bj}) and thus does not
deteriorate the approximation. However, when done in combination with a small
$\alphadip$ parameter~\cite{Nagy:1998bb}, enhanced finite contributions in
the subtracted and re-added real corrections are treated differently,
which increases the error of the DPA and  tends to worsen the
agreement with the full computation \cite{Denner:2020bcz}.
This can be cured in different ways.
One possibility is to calculate only the
$\alphadip$-independent part of the $I$-operators, \ie the one for $\alphadip=1$,
within the DPA but the $\alphadip$-dependent part, which contains
logarithms of $\alphadip$, with full kinematics.
This allows to cancel exactly the
$\alphadip$-dependence in the sum of integrated and subtraction dipoles.
Another option is to apply the DPA only to the virtual corrections
with IR singularities subtracted
via an appropriate choice of regularisation parameters.%
\footnote{In practice we discard the IR poles and set the parameter
  {\tt muir} of {\sc Collier} equal to the top-quark mass.}  Since the
virtual corrections are smaller than the contributions of the
integrated dipoles, as shown in \refse{se:fulvsdpa}, this provides
actually a better  approximation of the calculation with full matrix elements.

\subsection{Validation}\label{se:valid}
The calculation presented in this paper has been validated at several levels.

The correct functioning of the subtraction of IR singularities has
been checked performing the computation with different values of the
$\alphadip$ parameters that enter the calculation of integrated and
unintegrated subtraction dipoles \cite{Nagy:2003tz}.  Setting
$\alphadip < 1$ enables one to restrict the phase-space integration to
the singular regions, in order to avoid problems from sizeable
remappings of momenta in the Monte Carlo integration. 
Varying these parameters is equivalent to changing the
definition of the subtraction counterterms, but the sum of the
subtracted real and integrated dipoles remains independent of the
choice of $\alphadip$.

As a validation of the virtual contributions, we have compared the
results based on full matrix elements with the corresponding DPA
results, including factorisable and non-factorisable QCD corrections
(see \refse{dpa}). The agreement between the two results is
astonishingly good.  At the integrated level, the discrepancy between
the two calculations is of order $0.2\%$. A more detailed discussion on
this comparison is given in \refse{se:fulvsdpa}.

The specific final state that we consider features three charged
leptons with different flavours. However, in order to further check
the computations with full matrix elements, we have independently
reproduced some results that have been recently published in
\citere{Bevilacqua:2020pzy}, where the considered final state involves
two identical positrons and a muon. Employing the same final state, SM
input parameters, selection cuts ($\pt{b} > 25\GeV$), PDF set
(NNPDF3.0), as well as factorisation and renormalisation scale,
%($\mu_{\rR}=\mu_{\rF}=H_{\rT}/3$),
we obtain the following integrated cross-sections:
\begin{align}
\sigma_{\rm LO} ={}&
0.1147(1)^{+0.0304}_{-0.0224} \,\fb
\quad ({\textrm {\citere{Bevilacqua:2020pzy}}}:
0.1151^{+0.0305}_{-0.0225}\, \fb)\,, \nnb\\
\sigma_{\rm NLO} ={}&
0.1247(4)^{+0.0046}_{-0.0078} \,\fb
\quad ({\textrm {\citere{Bevilacqua:2020pzy}}}:
0.1244^{+0.0043}_{-0.0077} \,\fb)\,. \nnb
\end{align}
The only difference between the two calculations lies in the treatment
of resonant electroweak vector bosons,
which in \citere{Bevilacqua:2020pzy} is performed in the fixed-width scheme, while in our
calculation the complex-mass scheme is employed (see \refse{subsec:virt}).
The agreement between the two  calculations is excellent, both at LO and at NLO QCD.
This holds for the central values of the integrated
cross-sections and for the theoretical uncertainties given by scale variations.

\begin{figure}
  \centering
  \subfigure[$H_{\rT}$ variable for visible particles (excluding additional light jet at NLO).\label{htvis}]
            {\includegraphics[scale=0.36]{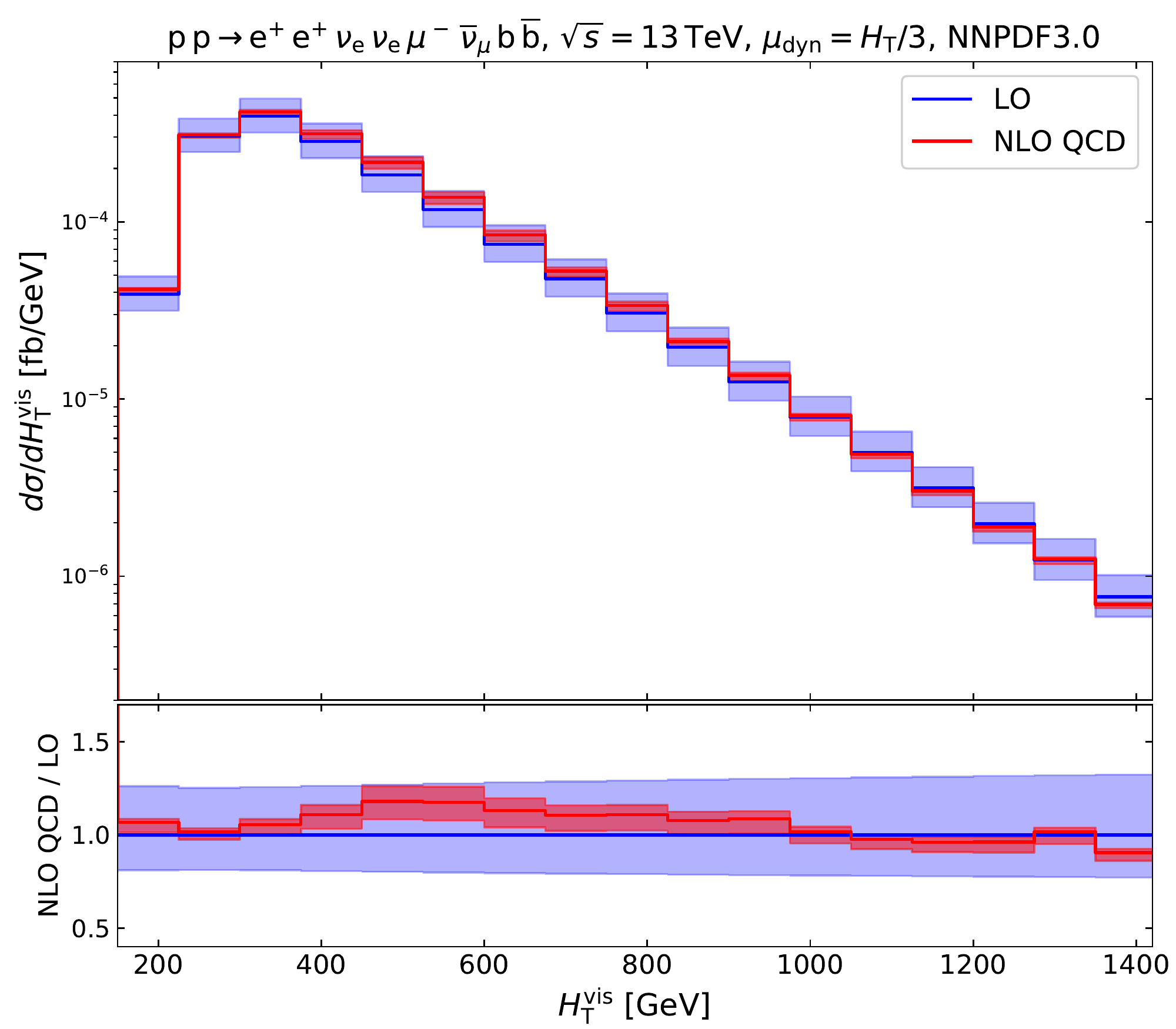}}
  \subfigure[Distance between the leading $p_{\rT}$ positron and the muon.\label{deltr}]
            {\includegraphics[scale=0.36]{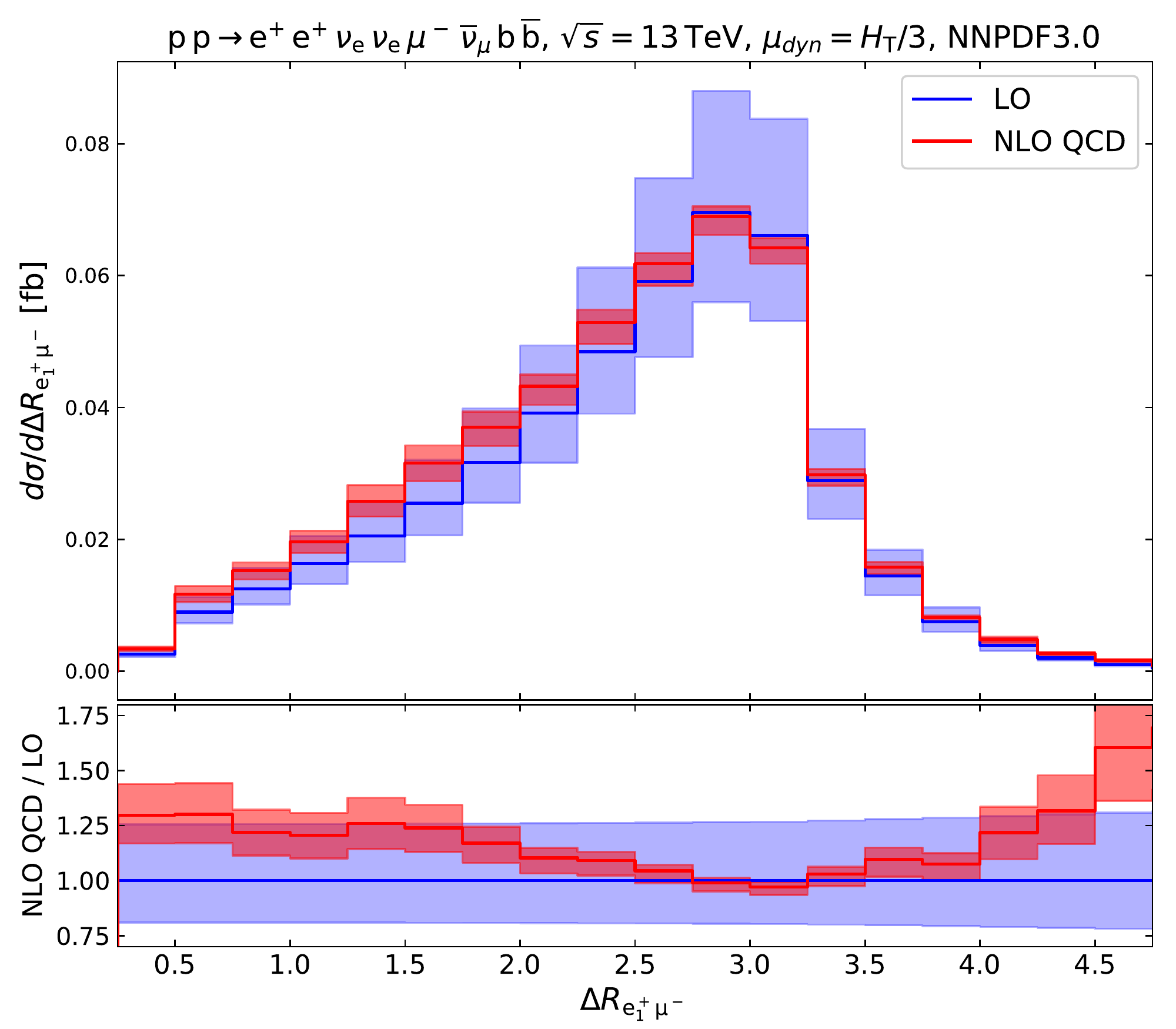}}
            \caption{Distributions at LO and NLO QCD for the process of \citere{Bevilacqua:2020pzy}:
              differential cross-sections (upper plot) and $K$-factor (lower plot).
              The uncertainty bands have been computed by means of 7-point variations of the
              factorisation and renormalisation scale around the central scale.}
            \label{fig:comparison}
\end{figure}

We have also studied some differential distributions in the setup of
\citere{Bevilacqua:2020pzy}, finding good agreement in the
absolute distributions and in the shapes of relative corrections. In
\reffi{fig:comparison} we show the LO and NLO distributions in the
variable $H_{\rT}^{\rm vis}$ that is the scalar sum of the transverse
momenta of all visible particles (excluding the additional radiated
light jet), and for the rapidity--azimuthal-angle distance between the
leading positron, \ie the positron with highest transverse momentum,
and the muon.
From a direct comparison with Figs.~5 and 7 of
\citere{Bevilacqua:2020pzy}, we find the same dependence of the
$K$-factors on the considered variables, as well as very similar
scale uncertainty bands in the various kinematic regions. We have
computed differential distributions in all of the kinematic variables
that are presented in \citere{Bevilacqua:2020pzy}, finding good
agreement in all cases.

\section{Numerical results}\label{sec:numresults}
\subsection{Input parameters}\label{sec:input}
In the following, we present results for the LHC at a CM energy of 13\TeV.  We consider
a final state with three charged leptons of different flavours, which are all
assumed to be massless, \ie also $m_\tau = 0$.
We note that our results can be used as well
for the case of identical leptons in the final state after applying
appropriate symmetry factor $1/2$ up to interference effects. These
interference effects are not doubly resonant and suppressed by factors
$\Gt/\Mt$. However, calculating the NLO corrections for identical
leptons does not pose any problems for our machinery and has in fact
been done to compare with \citere{Bevilacqua:2020pzy}, as described in
\refse{se:valid}.
We work in the five-flavour scheme and treat all light quarks,
including the $\Pb$~quarks, as massless.
A unit Cabibbo-Kobayashi-Maskawa matrix is understood.

%The masses and widths of the electroweak bosons 
%are chosen according to \citere{Tanabashi:2018oca}:
The on-shell values for the masses and widths of the electroweak bosons
are chosen according to \citere{Tanabashi:2018oca}:
\begin{align}
\Mwo ={}& 80.379 \GeV,\qquad\,\,\, \Gwo = 2.085\GeV\,, \nnb\\
\Mzo ={}& 91.1876 \GeV,\qquad \Gzo \,= 2.4952\GeV\,, \nnb\\
\MH  ={}& 125.00 \GeV, \qquad\,\,\, \GH\, \,\,= 0.00407\,\GeV\,.
\end{align}
The masses of the vector bosons are converted to their pole values by
means of the following relations \cite{Bardin:1988xt}:
\beq
\Mv = \frac{\Mvo}{\sqrt{1+({\Gvo}/{\Mvo})^2}}\,,\qquad\, 
\Gv = \frac{\Gvo}{\sqrt{1+({\Gvo}/{\Mvo})^2}}\,.
\eeq
The top-quark mass and widths are fixed as
\beq
\Mt = 173.0\GeV\,, \qquad \Gt^{\rm LO}   = 1.44\GeV\,, 
\qquad \Gt^{\rm NLO}   = 1.34\GeV\, .
\eeq
The top-quark width at LO has been computed based on the
formulas of \citere{Jezabek:1988iv} and using the pole mass and width
for the $\PW$ boson as input. The width at NLO QCD has been
obtained upon applying the relative NLO QCD corrections of
\citere{Basso:2015gca} to the LO width.

The electroweak coupling is
extracted from the Fermi constant $G_\mu$ \cite{Denner:2000bj} by
means of
\beq
\alpha = \frac{\sqrt{2}}{\pi}\,G_\mu\Mw^2\left[1-\left(\frac{\Mw^2}{\Mz^2}\right)\right]\,,
\eeq
where $G_\mu = 1.16638\cdot10^{-5} \GeV^{-2}$.

The masses of unstable particles, \ie the electroweak vector bosons
and the top quark are treated in the complex-mass scheme
\cite{Denner:1999gp,Denner:2000bj,Denner:2005fg,Denner:2006ic} in all
parts of the computation. As a consequence, the electroweak mixing
angle and the couplings become complex as well.

For the LO (NLO) calculation, we use NNPDF3.1 PDFs \cite{Ball:2014uwa}
computed at LO (NLO) with $\as=0.118$.  The strong coupling constant
$\as$ used in the calculation of the amplitudes matches the one used
in the evolution of PDFs.  The PDFs and the running of $\as$ are
obtained by interfacing \mocanlo with {\scshape LHAPDF6}
\cite{Buckley:2014ana}.

QCD partons with $|\eta|<5$ are clustered into jets by means of
the anti-$k_t$ algorithm \cite{Cacciari:2008gp} with resolution
radius $R=0.4$.

Our choice of selection cuts reflects those applied by ATLAS in a
recent analysis \cite{Aaboud:2019njj} (see Table 5 therein).  In
particular, we ask for exactly two $\Pb$~jets in the final state,
assuming a perfect $\Pb$-tagging efficiency, 
%(estimated to be $77\%$ from ${\Pt\overline{\Pt}}$ measurements). 
which are required to fulfil
\beq
\pt{\Pb} > 25 \GeV\,, \quad |\eta_{\Pb} |<2.5\,.
\eeq
We apply basic transverse-momentum, rapidity and isolation cuts to the
three charged leptons,
\beq
\pt{\Pl} > 27 \GeV\,, \quad |\eta_\Pl |<2.5\,, \quad \Delta R_{\Pl \Pb}>0.4\,.
\eeq
While these cuts are applied to the light leptons $\Pl=\Pe,\mu$ by
ATLAS, we apply them also to $\tau$~leptons. We remind the reader
that our results can be applied as well for final states with identical
leptons within a good approximation.  No specific veto is imposed on
the additional light jet that may come from real radiation at NLO QCD,
in case it is not recombined with a $\Pb$~jet.  Furthermore, we do not
impose any restriction on the missing transverse momentum.
%An additional cut $|M_{\Pl^+\Pl^-}-\Mz|>10\GeV$ would be required if two
%leptons with same flavour are considered. However, we consider in this paper
%the simple case of three charged leptons with different flavours.

\subsection{Fiducial cross-sections}\label{sec:integrated}
A common choice in the literature for the factorisation and
renormalisation scale for the considered process is \cite{deFlorian:2016spz,Garzelli:2012bn}
\beq\label{eq:scaleA}
\mu_0^{\rm (a)} = \Mt + \frac{\Mw}{2}\,,
\eeq
which is adapted to the masses of the leading resonance structure
contributing to the final state under investigation.  In addition to
this fixed scale, we have considered dynamical scales that depend on
the transverse momenta of the final-state particles.  One choice
depends on the $H_{\rT}$ variable, defined as
\beq\label{defHT}
H_{\rT} = \pt{\rm miss}+\sum_{i = {\Pb,\Pl}} \pt{i} \,,\nnb
\eeq
where the sum runs over all $\Pb$~jets and charged leptons, excluding
additional light jets that may arise from real radiation at NLO. We
have checked that the inclusion of the $p_{\rT}$ of additional
radiation would lead to a significant deterioration of the
scale dependence of the fiducial cross-section.  Since
$H_{\rT}$ is measurable at the LHC, it represents a
natural scale choice in the computation of the full off-shell process.  We
have investigated two different variants of scales based on
$H_{\rT}$,
\beq\label{eq:scaleBC}
\mu_0^{\rm (b)} = \frac{H_{\rT}}{2} \qquad {\rm and}\qquad \mu_0^{\rm (c)} = \frac{H_{\rT}}{3}\,,
\eeq
where the second one has been proposed in \citere{Bevilacqua:2020pzy}.

An alternative definition of a dynamical scale is based on the top and
antitop transverse masses, and is motivated by the choice made in
\citeres{Denner:2016wet, Denner:2017kzu} for $\Pt\overline{\Pt}$ and
$\Pt\overline{\Pt}\PH$ production at the LHC,
\beq\label{eq:scaleD}
\mu_0^{\rm (d)} = {\left(\tm{\Pt}\,\tm{\overline{\Pt}}\right)}^{1/2}=\left(\sqrt{\Mt^2+{\pt{\Pt}}^2}\,\sqrt{\Mt^2+{\pt{\overline{\Pt}}}^2}\right)^{1/2}  \,,
\eeq
where the top and antitop transverse momenta are reconstructed from
their decay products based on Monte Carlo truth. Note that the
determination of the top momentum is subject to an ambiguity, since it
can be reconstructed by two different lepton--neutrino pairs: we
choose the pair that, combined with the $\Pb$ quark, gives rise to an
invariant mass which is the closer to $\Mt$. This scale choice is
motivated by the expectation that top--antitop resonances dominate the
cross-section, even when considering complete off-shell effects.
Finally, we consider some results for the scale choice
\beq\label{eq:scaleE}
\mu_0^{\rm (e)} = \frac{1}{2}{\left(\tm{\Pt}\,\tm{\overline{\Pt}}\right)}^{1/2}\,.
\eeq

In \refta{tablesigma} we show the integrated cross-sections for the
five scale choices, complemented with the corresponding scale
uncertainties, evaluated with 7-point scale variations, namely
rescaling the central factorisation and renormalisation scale by the
factors
\beq
%\big\{
\left(0.5,0.5\right),\left(1,0.5\right),\left(0.5,1\right),\left(1,1\right),\left(1,2\right),\left(2,1\right),\left(2,2\right)\,.\nnb
%\big\}\,.
\eeq
When performing scale variations, the NLO QCD top-quark width is kept fixed.
\begin{table}
\begin{center}
\begin{tabular}{C{4.2cm}|C{3.7cm}C{3.7cm}C{1.6cm}}
\hline
\cellcolor{blue!9}  central scale & {\cellcolor{blue!9}  LO } & {\cellcolor{blue!9}  NLO QCD} & \cellcolor{blue!9}  $K$-factor\\
%\hline
\hline\\[-0.35cm]
  $\mu_0^{\rm (a)} = \Mt + {\Mw}/{2}$   &  $0.2042(1)^{+0.0485(23.8\%)}_{-0.0367(18.0\%)}$   &  $0.2452(7)^{+0.0109(4.5\%)}_{-0.0166(6.8\%)}$ & 1.20\\[0.3cm]
%\hline
%\hline
$\mu_0^{\rm (b)} = H_{\rT}/2$ &  $0.1931(1)^{+0.0444(23.0\%)}_{-0.0339(17.5\%)}$   &  $0.2330(9)^{+0.0098(4.2\%)}_{-0.0152(6.5\%)}$ & 1.21\\[0.3cm]
%\hline
$\mu_0^{\rm (c)} = H_{\rT}/3$  &  $0.2175(1) ^{+0.0525(24.2\%)}_{-0.0396(18.2\%)}$   &  $0.2462(8)^{+0.0069(2.8\%)}_{-0.0143(5.8\%)}$ & ${1.13}$\\[0.3cm]
$\mu_0^{\rm (d)} = {\left(\tm{\Pt}\,\tm{\bar{\Pt}}\right)}^{1/2}$  &  $0.1920(1)^{+0.0441(23.0\%)}_{-0.0336(17.5\%)}$   &  $0.2394(6)^{+0.0129(5.4\%)}_{-0.0172(7.2\%)}$ & 1.25\\[0.3cm]
$\mu_0^{\rm (e)} = {\left(\tm{\Pt}\,\tm{\bar{\Pt}}\right)}^{1/2}/2$  &  $0.2360(1)^{+0.0591(24.9\%)}_{-0.0441(18.7\%)}$   &  $0.2535(8)^{+0.0086(3.4\%)}_{-0.0133(5.2\%)}$ & 1.07\\[0.2cm]
\hline
\end{tabular}
\end{center}
\caption{Fiducial cross-sections (fb) for fixed and dynamical scale
  choices defined in this section. The uncertainties 
have been computed with 7-point scale variations around the central value. }
\label{tablesigma}
\end{table}%
We have checked that the largest scale variations result
from varying the renormalisation scale.  At LO the fiducial cross-sections
computed with $\mu_0^{\rm (b)}$ and $\mu_0^{\rm (d)}$ differ by less
than $1\%$ and are about $6\%$ smaller than the results with fixed scale
$\mu_0^{\rm (a)}$.  The scale $\mu_0^{\rm (c)}$ gives a cross-section
which is $6\%$ higher than the one obtained with the fixed scale,
while the one for  $\mu_0^{\rm (c)}$ is even $16\%$ larger.  In
all cases the scale uncertainties are of the same size, \ie
approximately $20\%$ relative to the central value.  The QCD
corrections amount to $+20\%$ for the fixed scale.  While the scale
$\mu_0^{\rm (b)}$ yields almost the same $K$-factor, the
corrections are larger for $\mu_0^{\rm (d)}$ ($+25\%$) and smaller for
$\mu_0^{\rm (c)}$ ($+13\%$)
and $\mu_0^{\rm (e)}$ ($+7\%$).
At NLO QCD, all scale choices give scale
uncertainties of the order of $5\%$ with some small
differences in the various cases. 
In particular, the use of $\mu_0^{\rm (c)}$ or $\mu_0^{\rm (e)}$
entails slightly smaller scale uncertainties compared to the
fixed scale, while those of $\mu_0^{\rm (d)}$ are larger. 
These differences are more sizeable at the differential level, as is
shown in the following.

\begin{figure}
  \centering
  \includegraphics[scale=0.36]{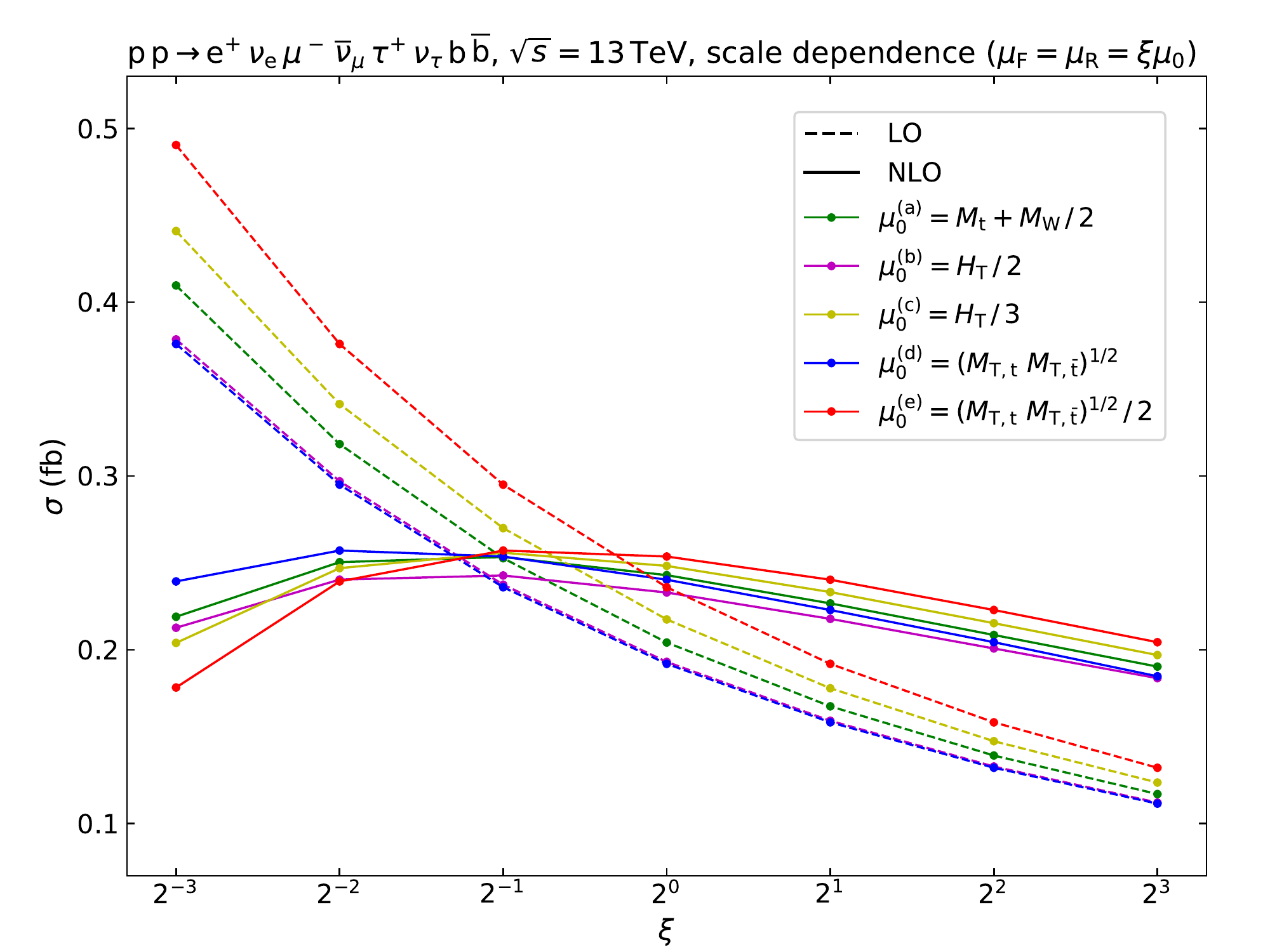}
  \caption{Scale dependence of the fiducial cross-section:
    factorisation and renormalisation scales are varied
    simultaneously about the central value $\mu_0$ with
    scale factor $\xi=\mu/\mu_0$.}\label{scaledepfig}
\end{figure}

In \reffi{scaledepfig} we show the scale dependence of the fiducial
cross-section for the five scale choices introduced in
Eqs.~\refeq{eq:scaleA}--\refeq{eq:scaleE}.
The factorisation and renormalisation scales are varied simultaneously
scaling the central value $\mu_0$ by a factor $\xi=1/8, 1/4, 1/2, 1,
2, 4, 8$.  At LO, $\mu_0^{\rm (b)}$ and $\mu_0^{\rm (d)}$ give almost
identical cross-sections for any $\xi$ factor, which are roughly $5\%$
smaller than the corresponding results with fixed scale $\mu_0^{\rm (a)}$.
On the contrary, the scale $\mu_0^{\rm (c)}$ gives a $+6\%$ enhancement to
the LO cross-sections with respect to the fixed scale.  At NLO QCD,
for $\xi>1$, the cross-section computed with $\mu_0^{\rm (a)}$ takes
intermediate values between the corresponding value for $\mu_0^{\rm (c)}$
($+7\%$) and for $\mu_0^{\rm (d)}$ ($-7\%$). The scale $\mu_0^{\rm (b)}$
behaves similarly to $\mu_0^{\rm (d)}$.  At small values of the scales,
$\xi<1$, the behaviour of the results for $\mu_0^{\rm (c)}$ and
$\mu_0^{\rm (d)}$ is inverted: for the former the cross-section becomes
noticeably lower than the one for fixed scale, while for the latter it
becomes larger.  Differently from $\mu_0^{\rm (d)}$, the results obtained
with $\mu_0^{\rm (b)}$ for $\xi < 1$ remain lower than those obtained with
$\mu_0^{\rm (a)}$.
The results for the central scale choice $\mu_0^{\rm (e)}=\mu_0^{\rm (d)}/2$
are close to those for $\mu_0^{\rm (c)}$, in particular for $\xi>1$. On the one hand,
such a choice reduces the corresponding scale dependence by giving
a flatter curve  around the central value. Furthermore, it shifts the
central scale towards the maximum of the curve in a similar way as
going from $\mu_0^{\rm (b)}$ to $\mu_0^{\rm (c)}$ and leads to a smaller
$K$-factor. On the other hand, $\mu_0^{\rm (d)}$ corresponds to the
dynamical choice made in calculations  for on-shell top quarks
\cite{Denner:2016wet, Denner:2017kzu}. 

\subsection{Differential cross-sections}\label{differential}

Beyond integrated cross-sections, it is of crucial importance to study
how NLO QCD corrections affect the distributions in the relevant
kinematic variables. This is the aim of this section.  We start by
comparing distributions obtained with different central-scale choices
in \refse{subsec:scalediff} and the corresponding scale variation of
the relative corrections. Thereafter we present more results for the
dynamical scale $\mu_0^{\rm (d)}$ in \refse{subsec:ourscale}.

\subsubsection{Comparison of different central-scale definitions}\label{subsec:scalediff}
Motivated by the results obtained at the integrated level, we are not
showing any result for $\mu_0^{\rm (b)}$. The focus is put on the
differences between the results obtained with the fixed scale and
those obtained with $\mu_0^{\rm (c)}$ and $\mu_0^{\rm (d)}$, as we
expect that the choice of a well-motivated dynamical scale is
beneficial for a better behaviour of NLO QCD corrections, in
particular, in the tails of energy-dependent distributions.

\begin{figure}[htb]
  \centering
  \subfigure[Fixed scale $\mu_0^{\rm (a)}=\Mt+\Mw/2$.\label{pte_fix}]{\includegraphics[scale=0.36]{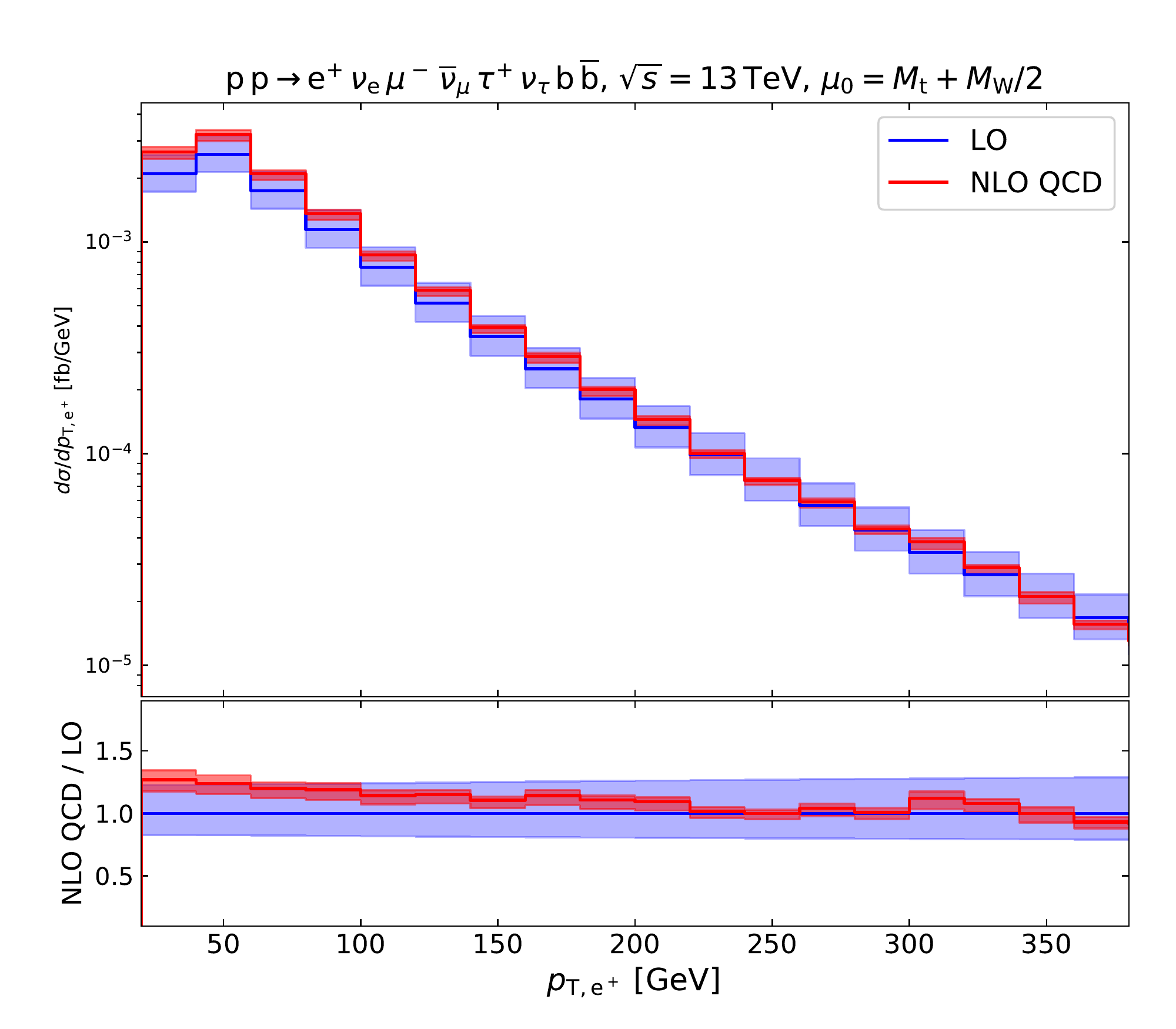}}
  \subfigure[Dynamical scale $\mu_0^{\rm (d)}={\left(\tm{\Pt}\,\tm{\bar{\Pt}}\right)}^{1/2}$.\label{pte_dyn}]{\includegraphics[scale=0.36]{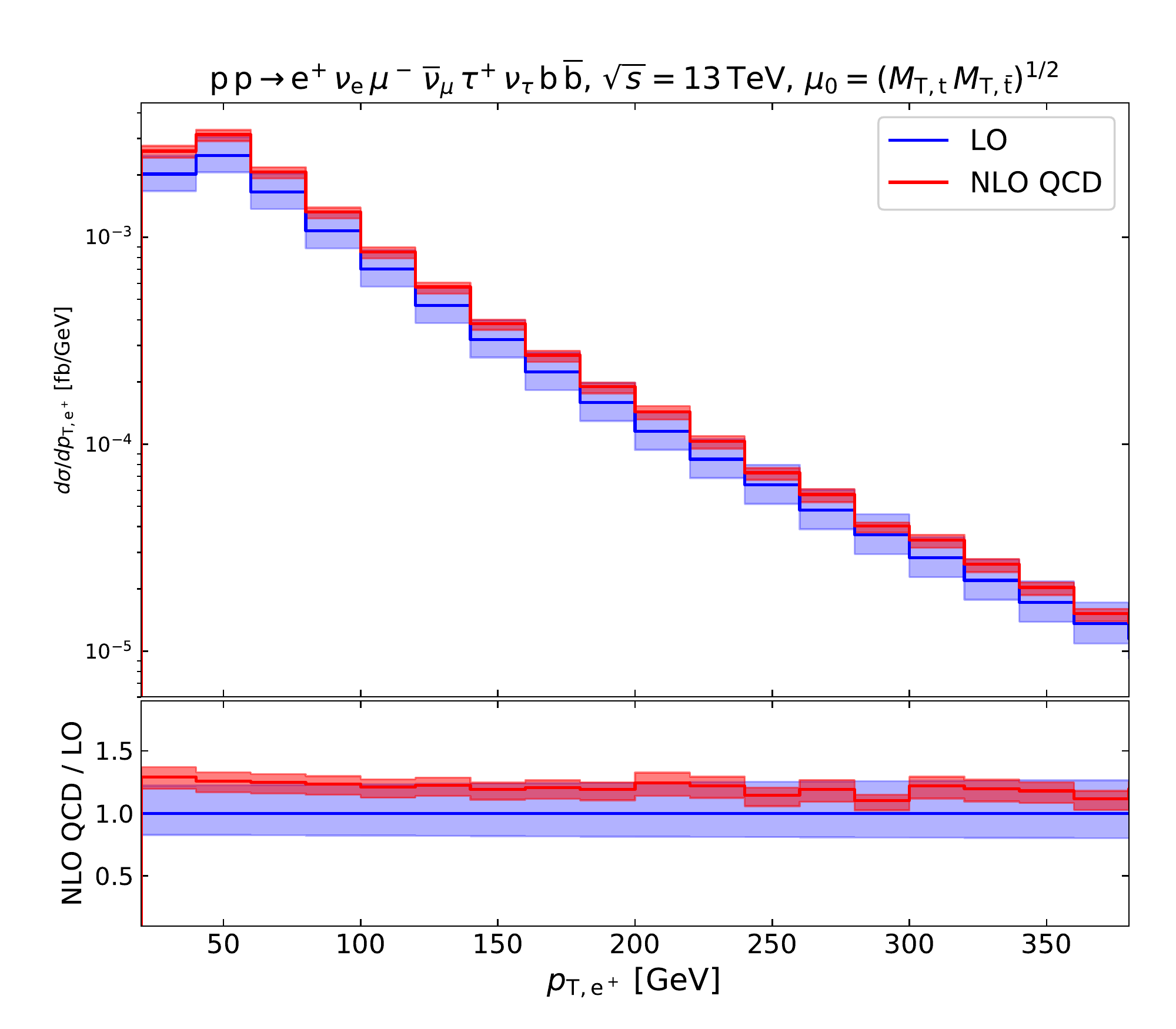}}
  \caption{Distributions in the transverse momentum of the positron at LO and NLO QCD: differential cross-sections (upper plot) and $K$-factor (lower plot). The uncertainty bands have been computed by means of 7-point variations of the factorisation and renormalisation scale around the central scale.
  }\label{pte}
\end{figure}

\begin{figure}[htb]
  \centering \subfigure[Fixed scale $\mu_0^{\rm
    (a)}=\Mt+\Mw/2$.\label{pttop_fix}]{\includegraphics[scale=0.36]{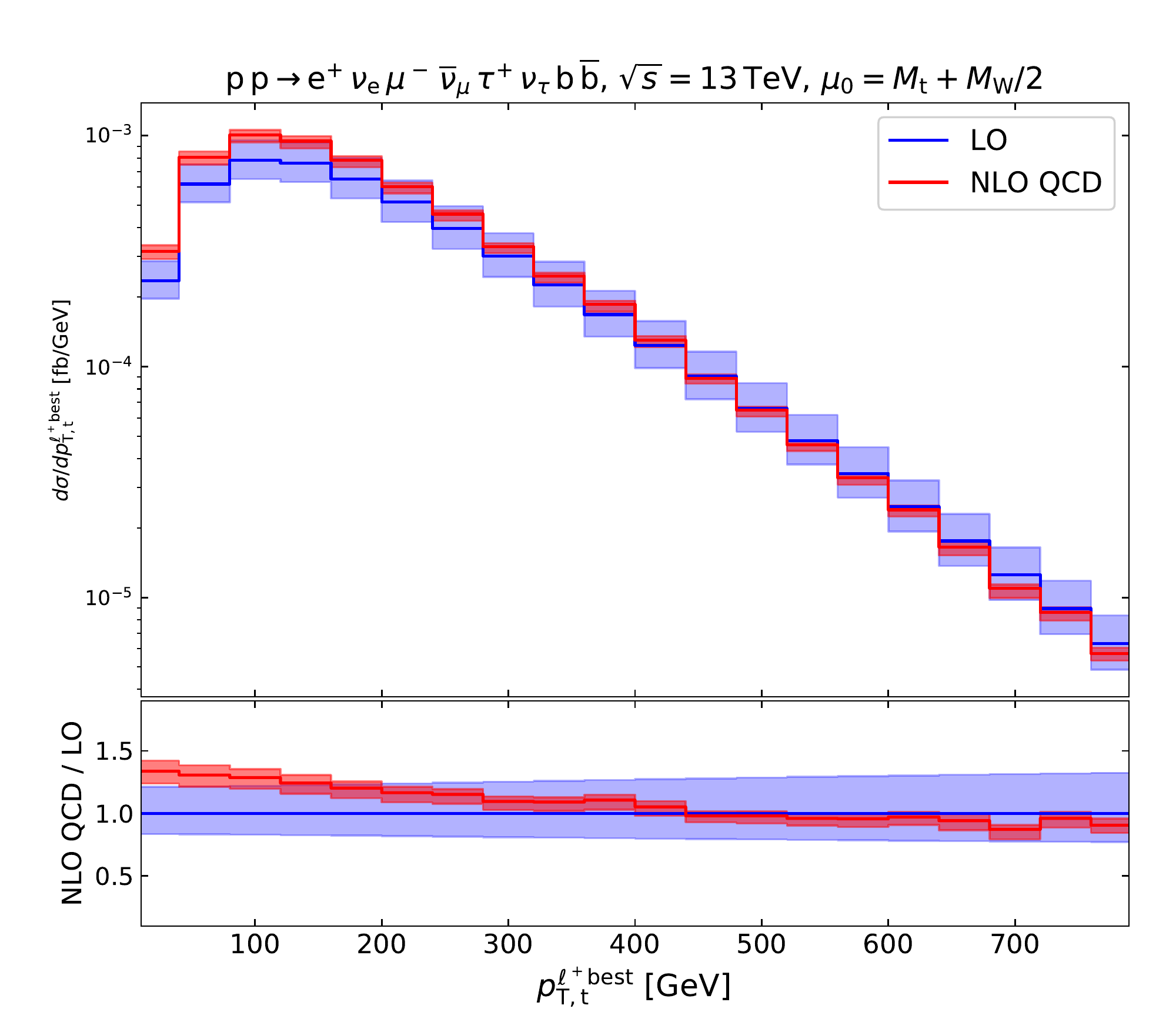}}
  \subfigure[Dynamical scale $\mu_0^{\rm
    (d)}={\left(\tm{\Pt}\,\tm{\bar{\Pt}}\right)}^{1/2}$.\label{pttop_dyn}]{\includegraphics[scale=0.36]{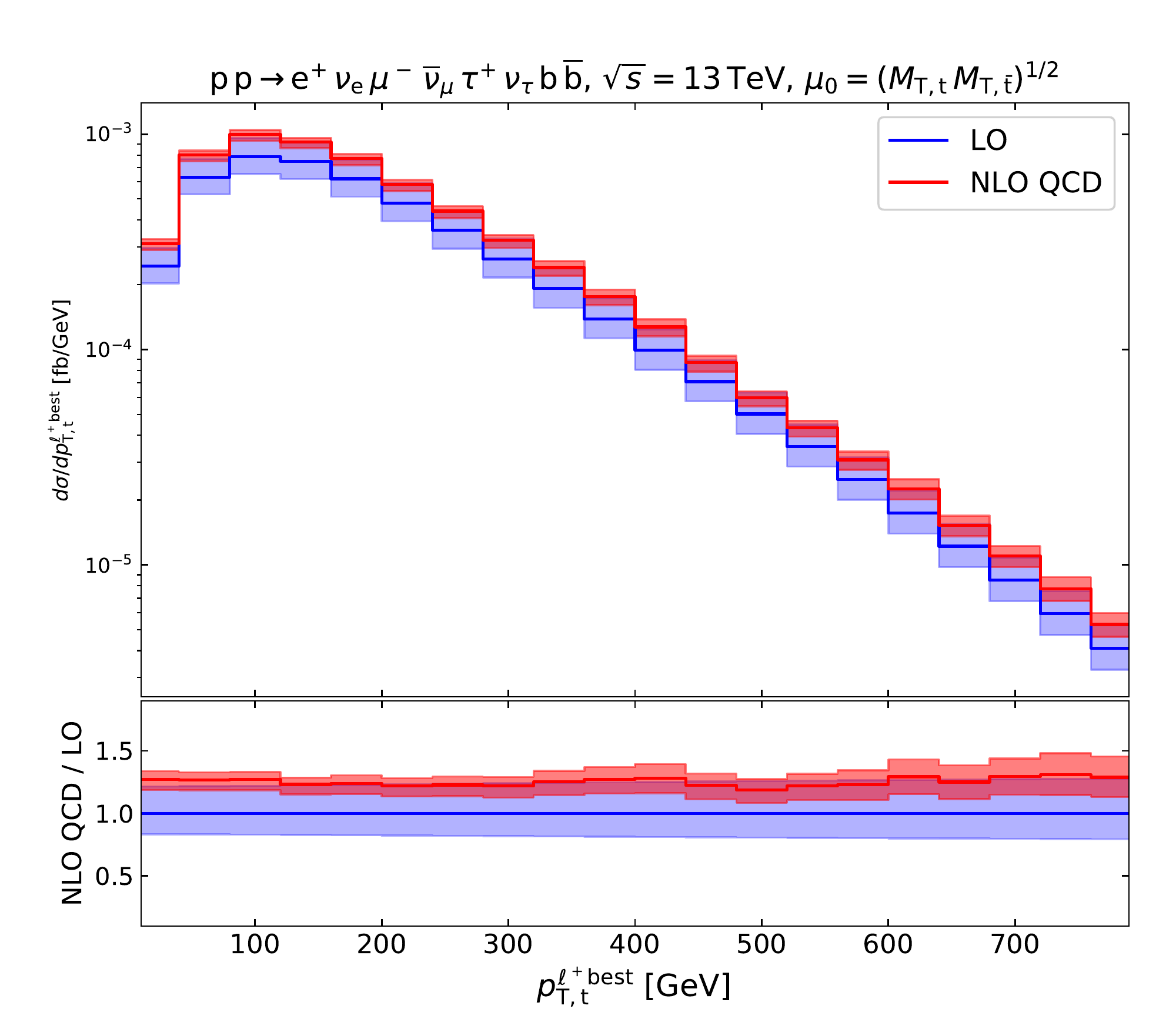}}
  \caption{Distributions in the transverse momentum of the
    reconstructed top quark at LO and NLO QCD: differential
    cross-sections (upper plot) and $K$-factor (lower plot). The
    uncertainty bands have been computed by means of 7-point
    variations of the factorisation and renormalisation scale around
    the central scale.
  }\label{pttop} 
\end{figure}

This is indeed the case for the transverse-momentum distributions for
the positron and the reconstructed top quark, which are shown
 in \reffi{pte} and \reffi{pttop}, respectively.
%The final state leptons kinematics is affected by NLO corrections
%as the additional radiation transverse momentum is not vetoed.
%
Comparing \reffi{pte_fix} with \reffi{pte_dyn}, it can be seen that
using the fixed scale leads to a differential $K$-factor that
drops monotonically from
%$1.27$ to $0.93$ 
$1.25$ to $0.95$ 
with increasing transverse momentum of the
positron.  On the contrary, the dynamical scale gives a $K$-factor that
decreases only from 
%$1.29$ to $1.12$.
$1.3$ to $1.1$.
Differently from the case of fixed scale, the scale uncertainties do
not decrease for the dynamical scale in the tails of the distribution.
Using the scale $\mu_0^{\rm (c)}$, the $K$-factor becomes somewhat  
flatter, but the scale uncertainties are similar.

The same situation can be found in the comparison of \reffi{pttop_fix}
and \reffi{pttop_dyn}.  The transverse momentum of the top quark
considered in \reffi{pttop} refers to the top quark reconstructed
using Monte Carlo truth from the bottom quark and the $\Pl^+\nu_\Pl$
pair that gives an invariant mass which is the closest to $\Mt$
(denoted ``$\Pl^+$ best'' in the following).  While not measurable at
the LHC, it is interesting to investigate this observable as it is
directly related to the dominant resonant structure of the process,
\ie the top and antitop quarks.  While the $K$-factor for
the fixed scale drops from 
%$1.34$ to $0.91$,
$1.35$ to $0.9$,
it is practically flat for
the  dynamical scale $\mu_0^{\rm (d)}$.  While the NLO
uncertainty bands are similar in the two cases in the soft part of the
spectrum they are sizeably larger for the dynamical scale at large
transverse momentum.  Again, the results are similar for the scale
$\mu_0^{\rm (c)}$, where the $K$-factor varies from $1.20$ to $1.12$. 

Given the flatness of the $K$-factors of \reffis{pte} and \ref{pttop},
we can conclude that using the dynamical scale $\mu_0^{\rm (d)}$ or
$\mu_0^{\rm (c)}$ results in better behaved NLO QCD corrections than
with the fixed scale.

\begin{figure}
  \centering
  \subfigure[Dynamical scale $\mu_0^{\rm (d)}={\left(\tm{\Pt}\,\tm{\bar{\Pt}}\right)}^{1/2}$\label{htvardyn}]{\includegraphics[scale=0.36]{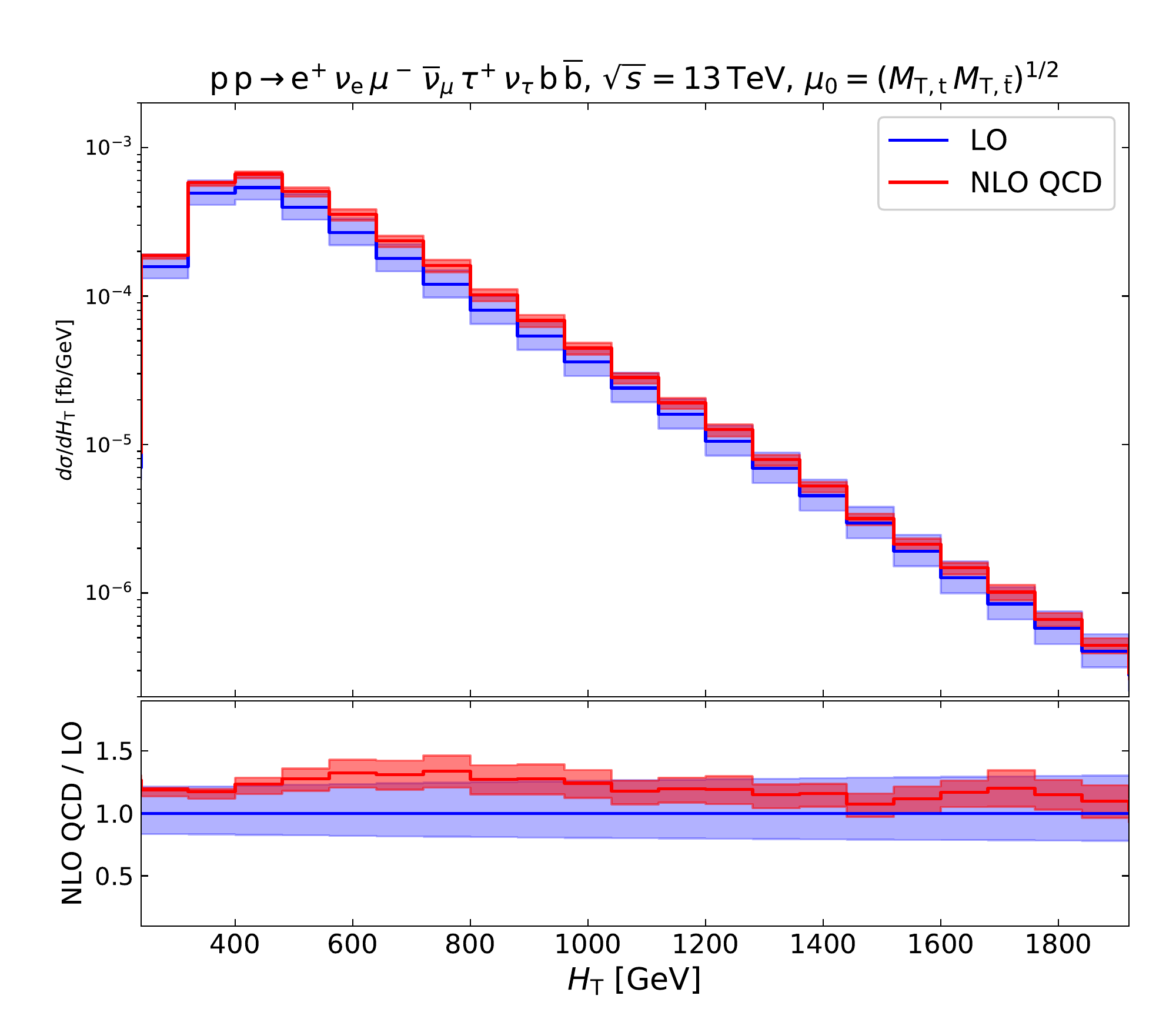}}
  \subfigure[Dynamical scale $\mu_0^{\rm (c)}=H_{\rT}/3$\label{htvarht3}]{\includegraphics[scale=0.36]{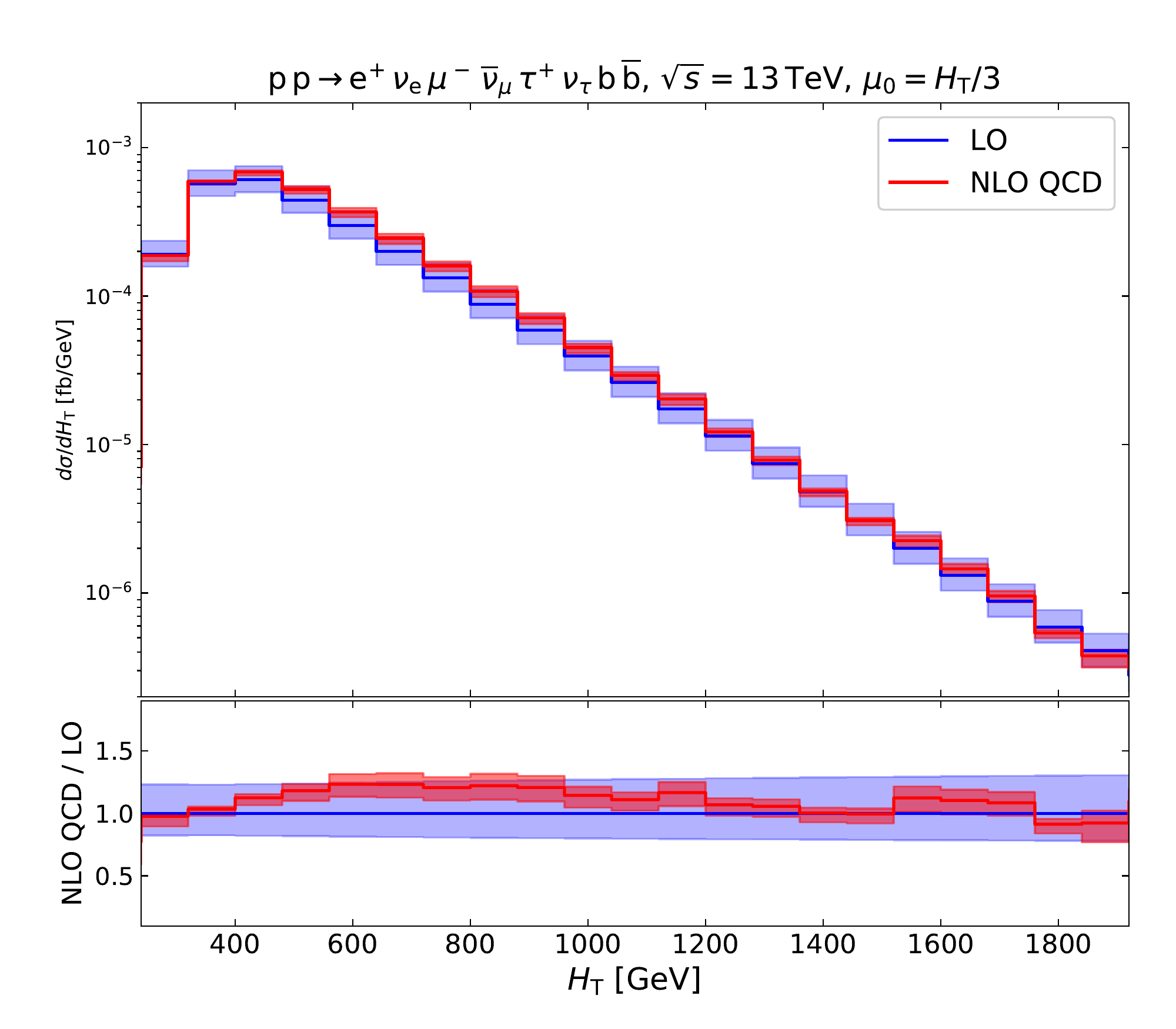}}  
  \caption{Distributions in the $H_{\rT}$ variable at LO and NLO QCD:
    differential cross-sections (upper plot) and $K$-factor (lower
    plot). The uncertainty bands have been computed by 
    7-point variations of the factorisation and renormalisation scale
    around the central scale. 
  }\label{htnorad}
\end{figure}

In \reffi{htnorad} we consider the distribution in the $H_{\rT}$
variable  comparing the results obtained with the two dynamical
scales $\mu_0^{\rm (d)}$ and $\mu_0^{\rm (c)}$.
The dynamical scale $\mu_0^{\rm (d)}$, which depends on the
top--antitop transverse masses, describes very well the $H_{\rT}$
variable at NLO QCD, with a $K$-factor varying between 
%$1.07$ and $1.33$.
$1.1$ and $1.35$.
Furthermore, the $K$-factor has a similar shape and variation
as the one obtained with $\mu_0^{\rm (c)}$, which coincides with the
variable itself up to the 1/3 multiplicative factor.  The most
relevant difference between the two scales is given by a constant
shift of the $K$-factor to larger values in the $\mu_0^{\rm (d)}$ case
resulting from the difference of the integrated cross-section (see
\refta{tablesigma}). The NLO uncertainty bands are marginally larger for
$\mu_0^{\rm (d)}$ for moderate values of $H_{\rT}$.
We have checked numerically that the uncertainty bands for $\mu_0^{\rm (e)}$ are 
comparable to those for $\mu_0^{\rm (c)}$.

Since analogous similarities between the two scale choices can be
found in the distributions for many other kinematic variables that we
have investigated, we conclude that the quality of the NLO
description of the process at the differential level is comparable
for the two dynamical scale choices. Therefore, in all results we are
going to present in the following, $\mu_0^{\rm (d)}$ is understood as
a central factorisation and renormalisation scale.

\subsubsection{Differential distributions for the default scale}\label{subsec:ourscale}% $\mu_0^{\rm (d)}$
In this section we discuss LO and NLO QCD results for selected
differential cross-sections, including both angular and
energy-dependent variables. We consider not only observables that are
directly measurable at the LHC, but also kinematic variables related
to the top and antitop resonances based on Monte Carlo truth.  The
explicit final state that we consider enables to sort charged leptons
by their flavour, with no need of a transverse-momentum ordering.
However, in order to be useful for final states with identical
leptons, we also compute some observables based on the
transverse-momentum ordering of the positively-charged leptons, as
needed in the case of identical positrons \cite{Bevilacqua:2020pzy}.
We denote leptons or bottom quarks with highest transverse
momenta as leading ones in the following.
Moreover, we also sort the two positively-charged leptons $\Pe^+,
\tau^+$ depending on how well they reconstruct the true top-quark
invariant mass, as already done in the previous sections for the
computation of the $\mu_0^{\rm (d)}$ central scale.

% azimuthal separations 
\begin{figure}
  \centering
  \subfigure[Azimuthal separation between the leading $\Pb$~jet
    and the leading positively-charged lepton.\label{dphi_bl_dyn}]{\includegraphics[scale=0.36]{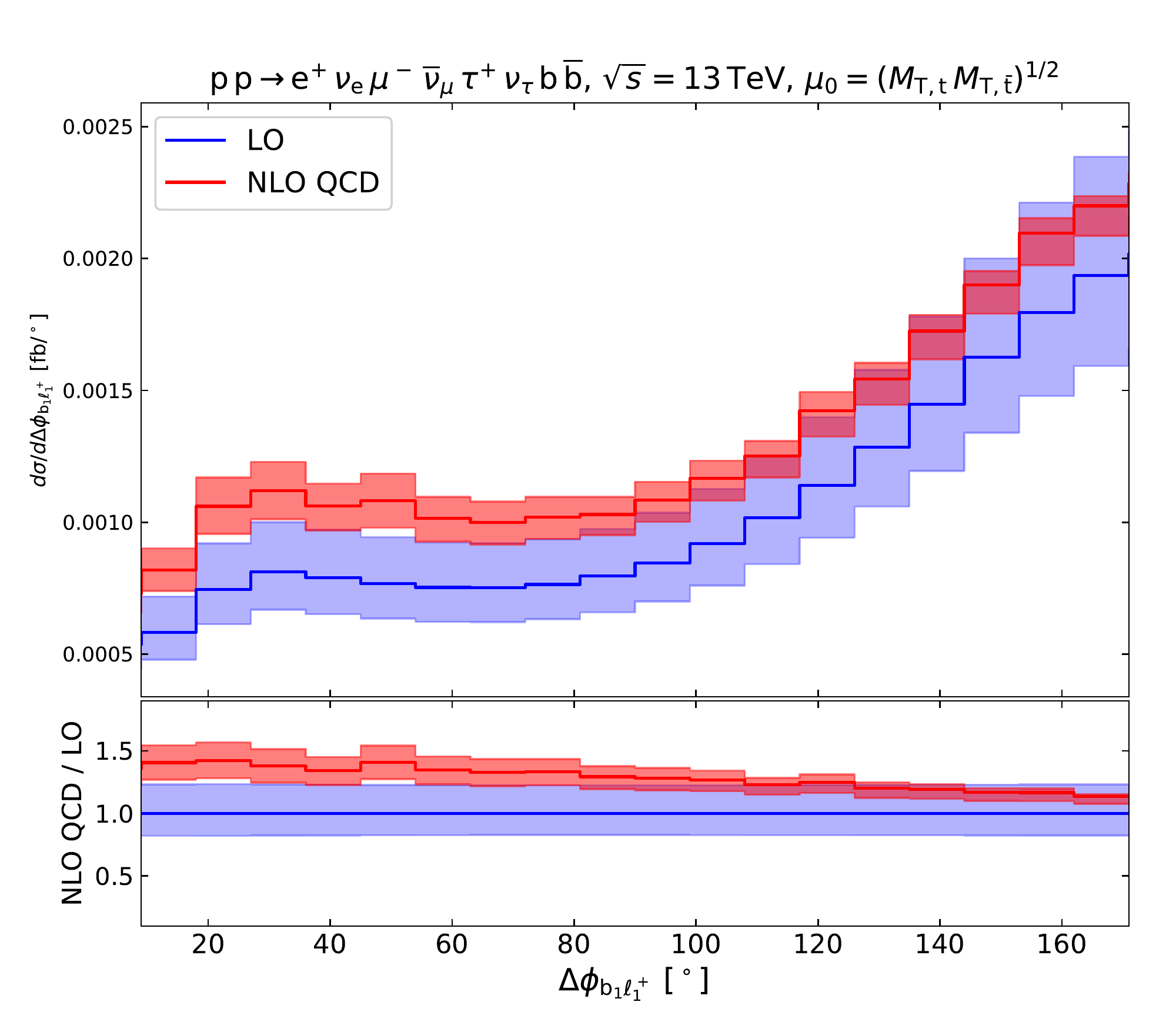}}
  \subfigure[Azimuthal separation between the positron
    and the muon.\label{dphi_emu_dyn}]{\includegraphics[scale=0.36]{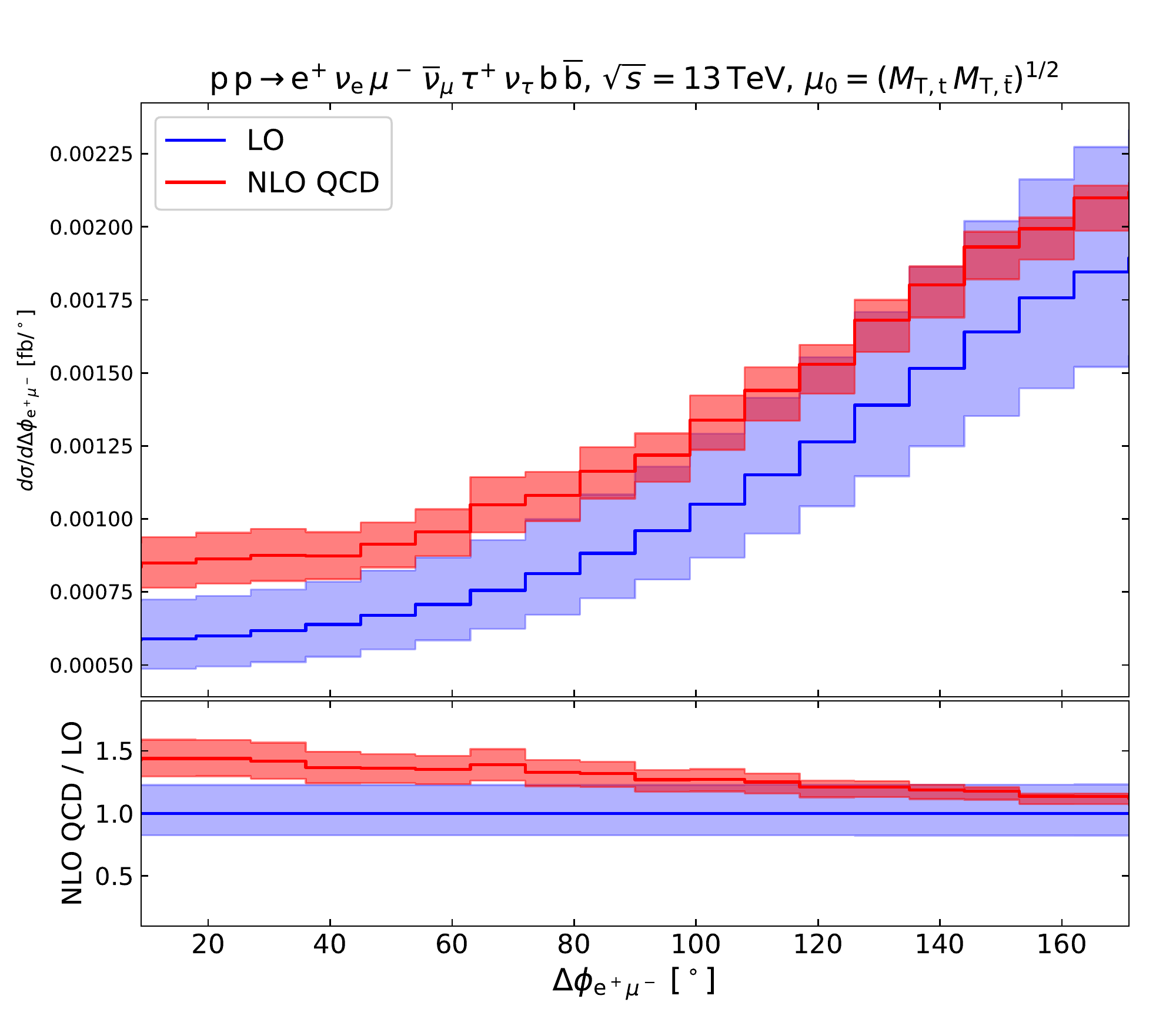}}
  \caption{Distributions at LO and NLO QCD for the dynamical scale choice:
    differential cross-sections (upper plot) and $K$-factor (lower plot).
    The uncertainty bands have been computed by means of 7-point
    variations of the factorisation and renormalisation scale around the central scale.}\label{dphi}
\end{figure}
In \reffi{dphi} we show the LO and NLO distributions and $K$-factors
for the azimuthal-angle separations between the leading $\Pb$~jet and
the leading positively-charged lepton [both sorted by their $p_{\rT}$,
\reffi{dphi_bl_dyn}] and between the positron and the muon
[\reffi{dphi_emu_dyn}].  The relative corrections for the two
distributions behave similarly. In fact, the $K$-factors are of order
$50\%$ in the region with smaller cross-section ($\Delta\phi \approx 0$)
and decrease to $15\%$ towards $\Delta\phi = \pi$, where both
distributions show a maximum.  While the NLO uncertainty bands are
almost as wide as the LO ones at low azimuthal separation, they become
noticeably smaller in the peak region.
We note that the distributions in the azimuthal separation between the
antitau and the muon (which are not shown here) have exactly the same
shape and normalisations as those shown in \reffi{dphi_emu_dyn}, since
the leptonic decay of the $\PW^+$~boson is universal and the symmetry
under the exchange of the two leptonically decaying $\PW^+$~bosons
is accounted for in the full amplitudes. This comment holds for any
variable that depends on the kinematics of a single positively-charged
lepton selected by its flavour.

\begin{figure}
  \centering
  \subfigure[Cosine of the angle between the two positively-charged leptons.\label{costh_dyn}]{\includegraphics[scale=0.36]{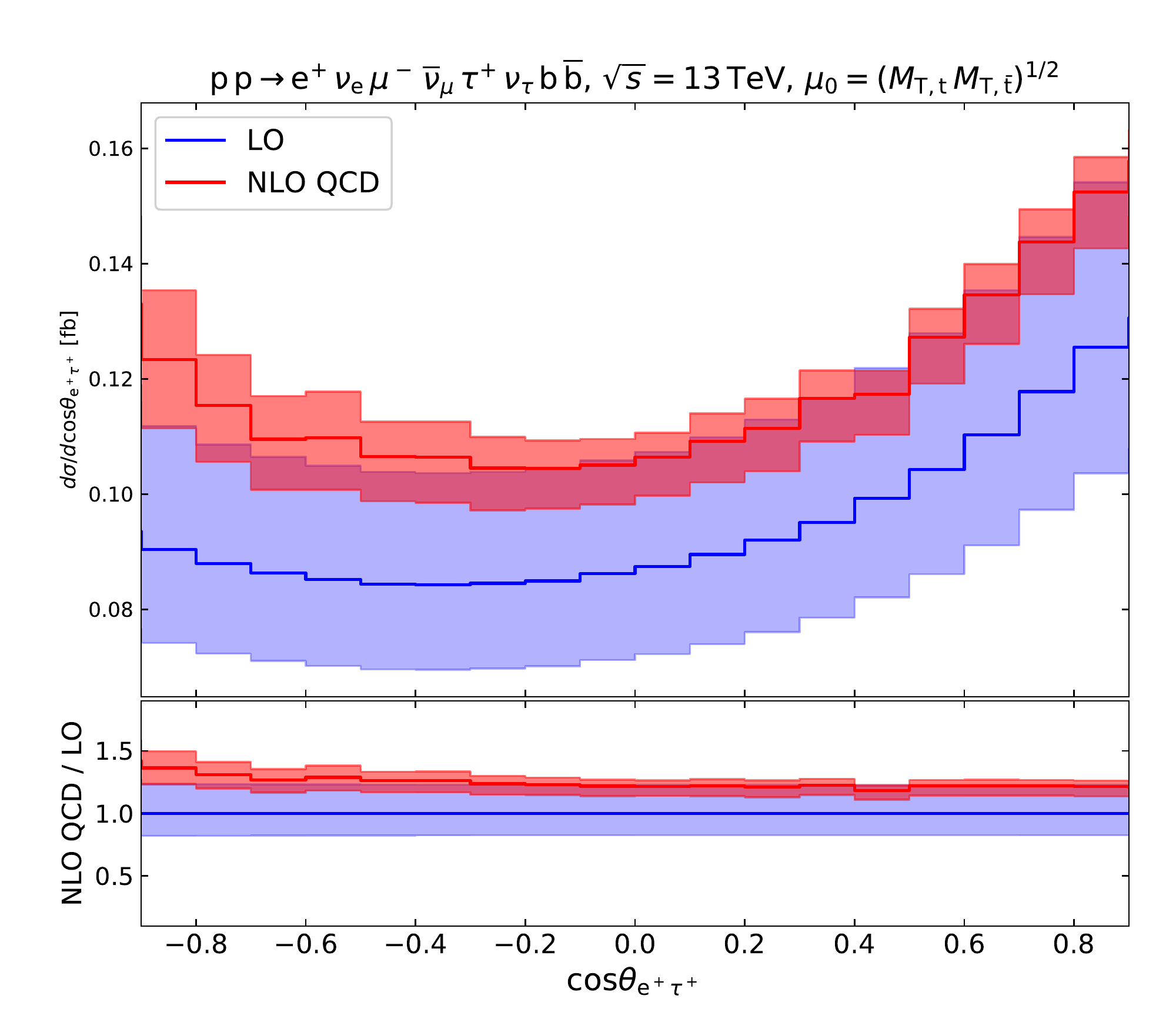}} 
  \subfigure[Invariant mass of the system formed by the positron and the leading $\Pb$~jet.\label{mbl_dyn}]{\includegraphics[scale=0.36]{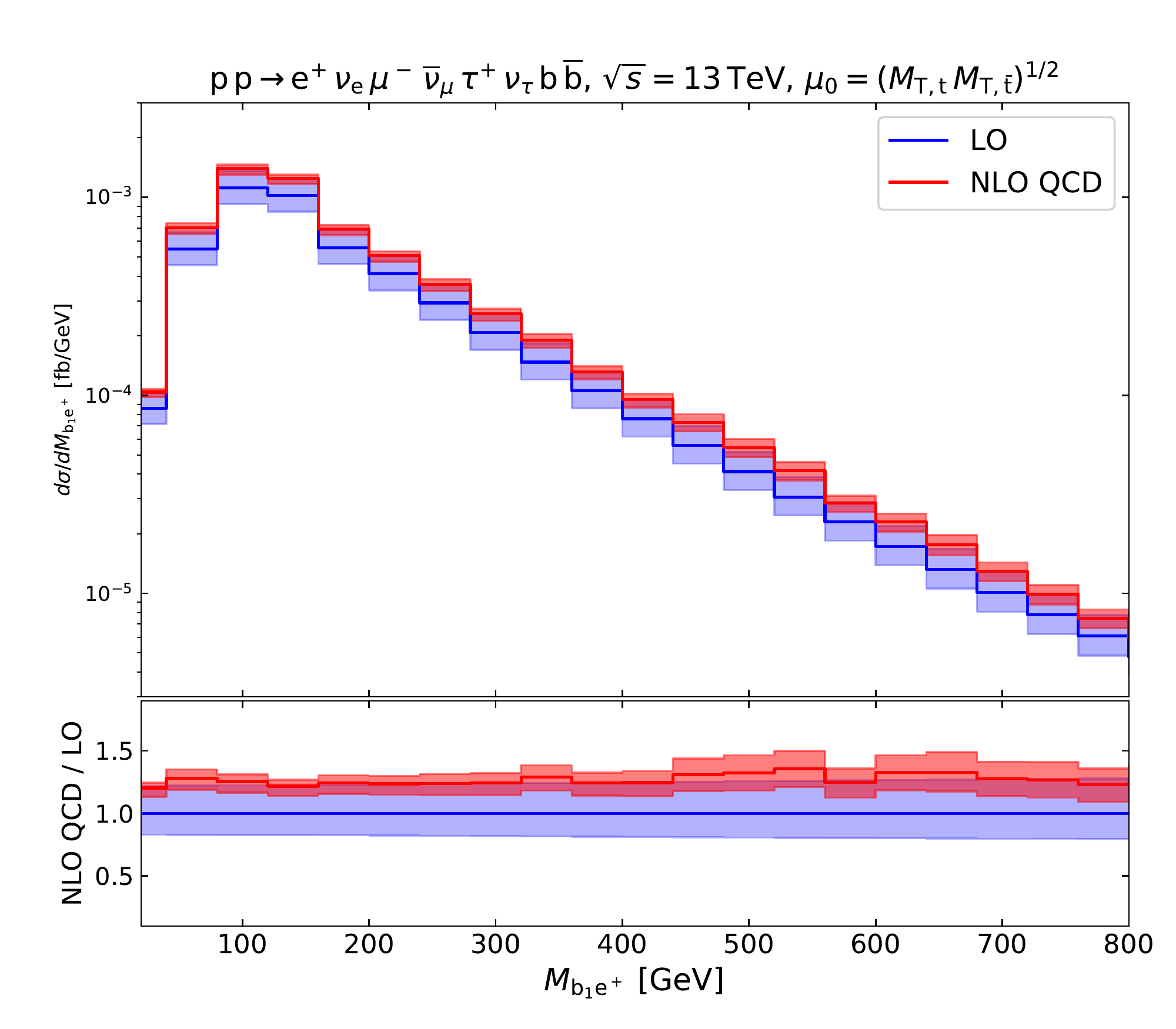}}   
  \caption{Same as in \reffi{dphi}.}\label{othervar}
\end{figure}
A slightly different situation is found for the distribution in the
cosine of the angle between the two positively charged
leptons, shown in \reffi{costh_dyn}.  This variable correlates the two
charged leptons which are decay products of the two $\PW^+$ bosons, \ie one of
them comes from the top-quark decay and the other one from the
$\PW^+$~boson recoiling against the $\Pt\overline{\Pt}$ system.  The
NLO corrections are positive over the whole spectrum, and inherit the
relative increase in the cross-section from the integrated results
($+25\%$). The $K$-factor 
decreases from 
%$1.42$ in the back-to-back configuration to $1.22$ 
$1.4$ in the back-to-back configuration to $1.2$ 
in
the collinear case. The minimum of the NLO distribution is shifted
from $\cos\theta_{\Pe^+\tau^+}\approx -0.35$ (at LO) to $ -0.2$, and
the two local maxima are in the collinear and anticollinear
configurations: the former configuration is preferred, even though NLO
corrections are larger for the latter.

In \reffi{mbl_dyn} we consider the distribution in the invariant mass
of the system formed by the positron and the leading $\Pb$~jet.  The
distribution exhibits a maximum around $100\GeV$ and a marked drop
around $150\GeV$. For on-shell W~boson and top quark, this variable
has an upper bound at $M_{\Pe^+\Pb}^2 = \Mt^2-\Mw^2$.
As the considered variable depends on the leading $\Pb$~jet, which is
not necessarily the bottom quark that originates from the top quark,
the drop is not as pronounced as it would be in the absence of this
ambiguity.  Nevertheless, the edge is present in the considered
process and could help to increase the experimental sensitivity to the
top-quark mass.  In the on-shell region, $M_{\Pe^+\Pb}<150\GeV$, the
NLO QCD corrections mildly decrease from $30\%$ to $25\%$, while in
the off-shell region, $M_{\Pe^+\Pb}>150\GeV$, they basically reflect
the overall normalisation with an almost flat $K$-factor, up to small
statistical oscillations in the tails of the distribution where the
rate is exponentially damped. The NLO uncertainty bands are of order $5\%$
in the part of the spectrum with a higher rate, while they increase
to $15\%$ in the tails.

\begin{figure}
  \centering
  \subfigure[Rapidity of the positron.\label{rappos_dyn}]{\includegraphics[scale=0.36]{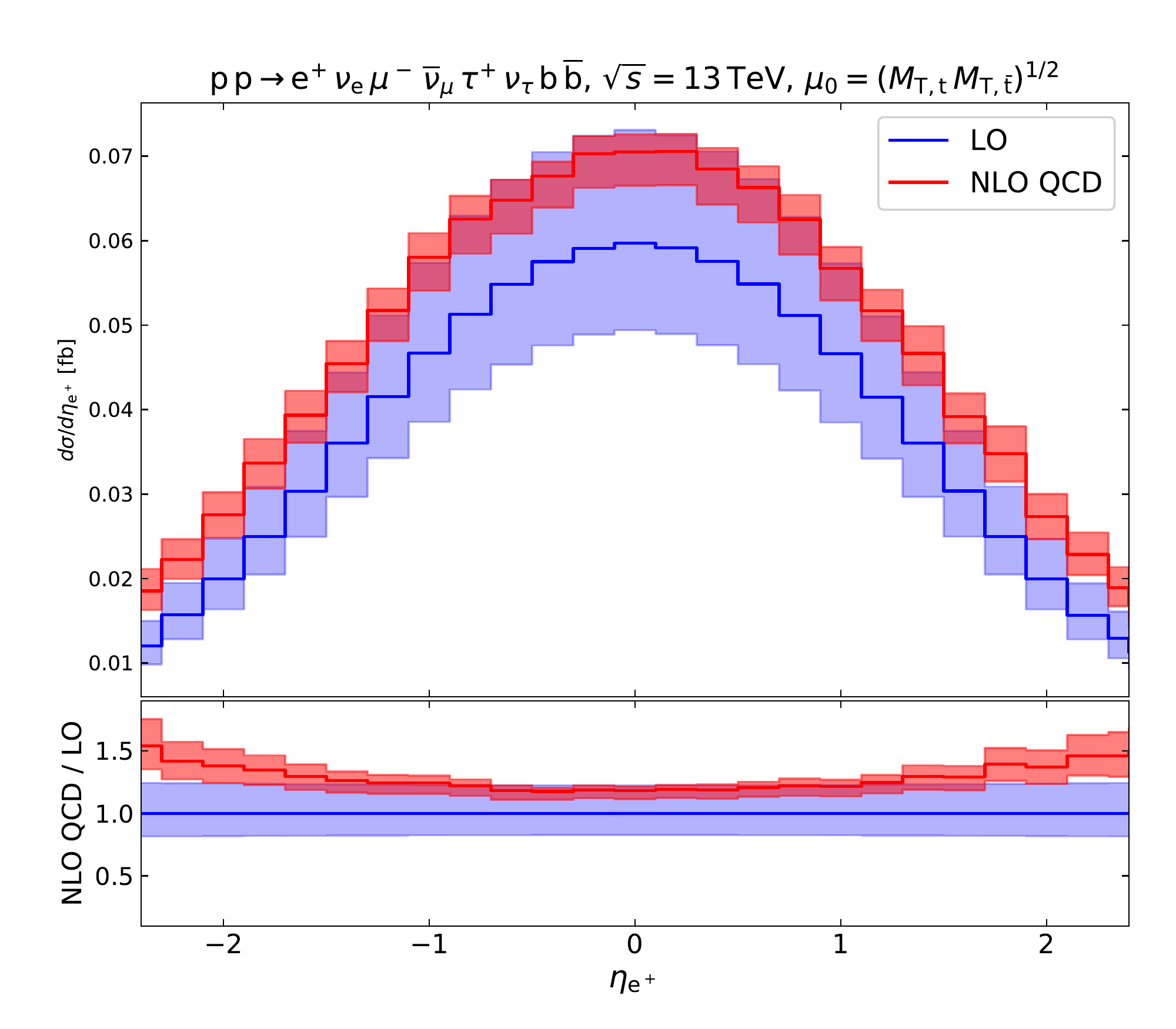}}
  \subfigure[Rapidity of the antitop quark.\label{raptop_dyn}]{\includegraphics[scale=0.36]{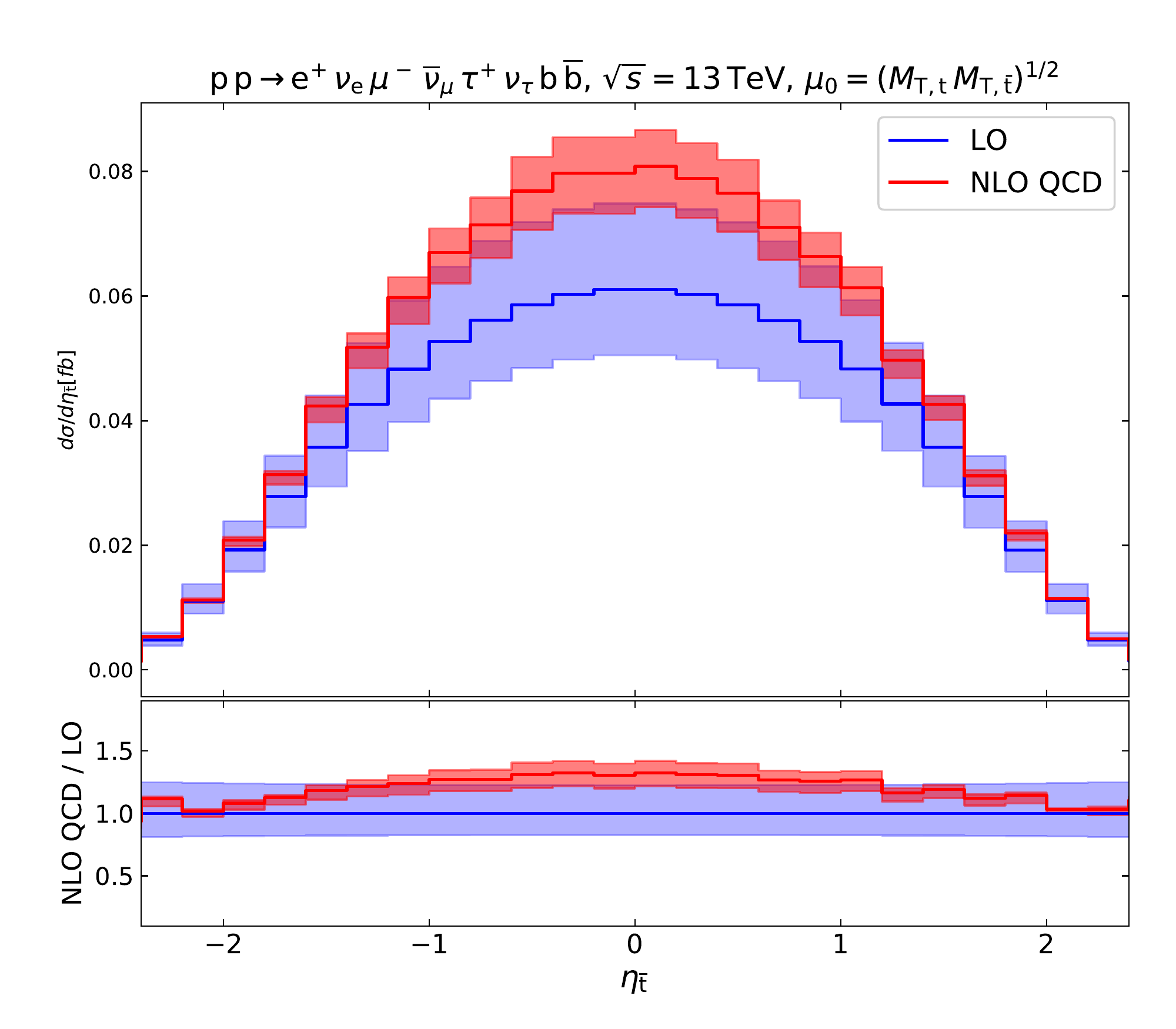}}
  \caption{Same as in \reffi{dphi}.}\label{rap}
\end{figure}
In \reffi{rap} we present the distributions in the rapidities of the
positron and the antitop quark.  The first distribution is measurable,
while the second one is based on Monte Carlo truth. The positron
rapidity is directly cut at $y_{\Pe^+}=\pm 2.5$, while the one of the
antitop gives an almost negligible contribution for $|y_{\bar{\Pt}}|>
2.5$.  Despite having similar behaviours at LO, the effect of QCD
radiative corrections is different in the two variables.  In the
positron case, the radiative corrections are maximal in the region
close to the acceptance cut (roughly $+50\%$), while they are smaller
near the peak of the distribution ($+20\%$ for $y_{\Pe^+}=0$).  In
contrast, the radiative corrections give a maximal enhancement to the
antitop distribution around the peak at zero~rapidity ($+35\%$), while
they are almost zero at large rapidities.

\begin{figure}
  \centering
  \subfigure[Invariant mass of the antitop quark.\label{invantitop_dyn}]{\includegraphics[scale=0.36]{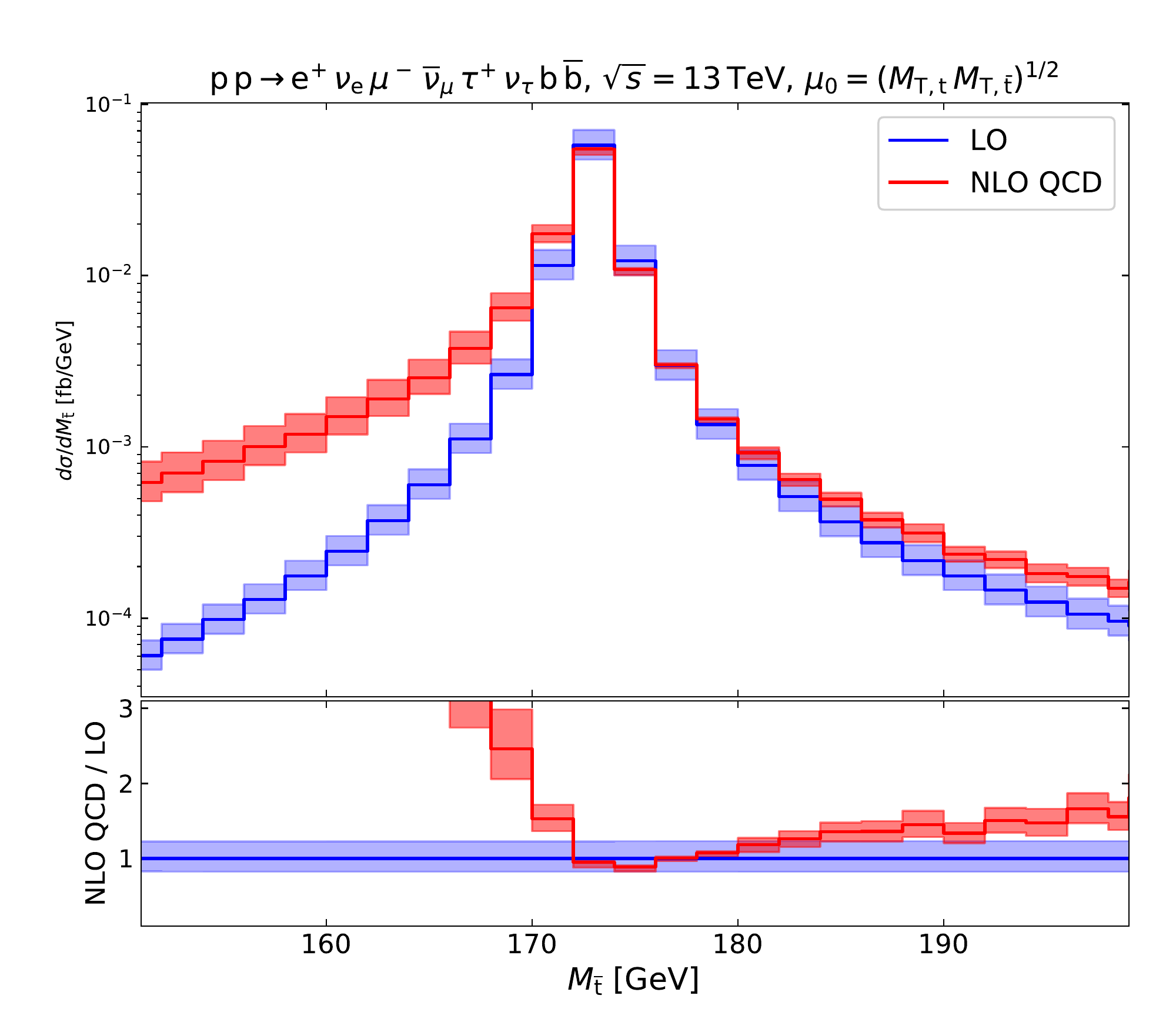}}
  \subfigure[Invariant mass of the reconstructed top quark.\label{invtop_dyn}]{\includegraphics[scale=0.36]{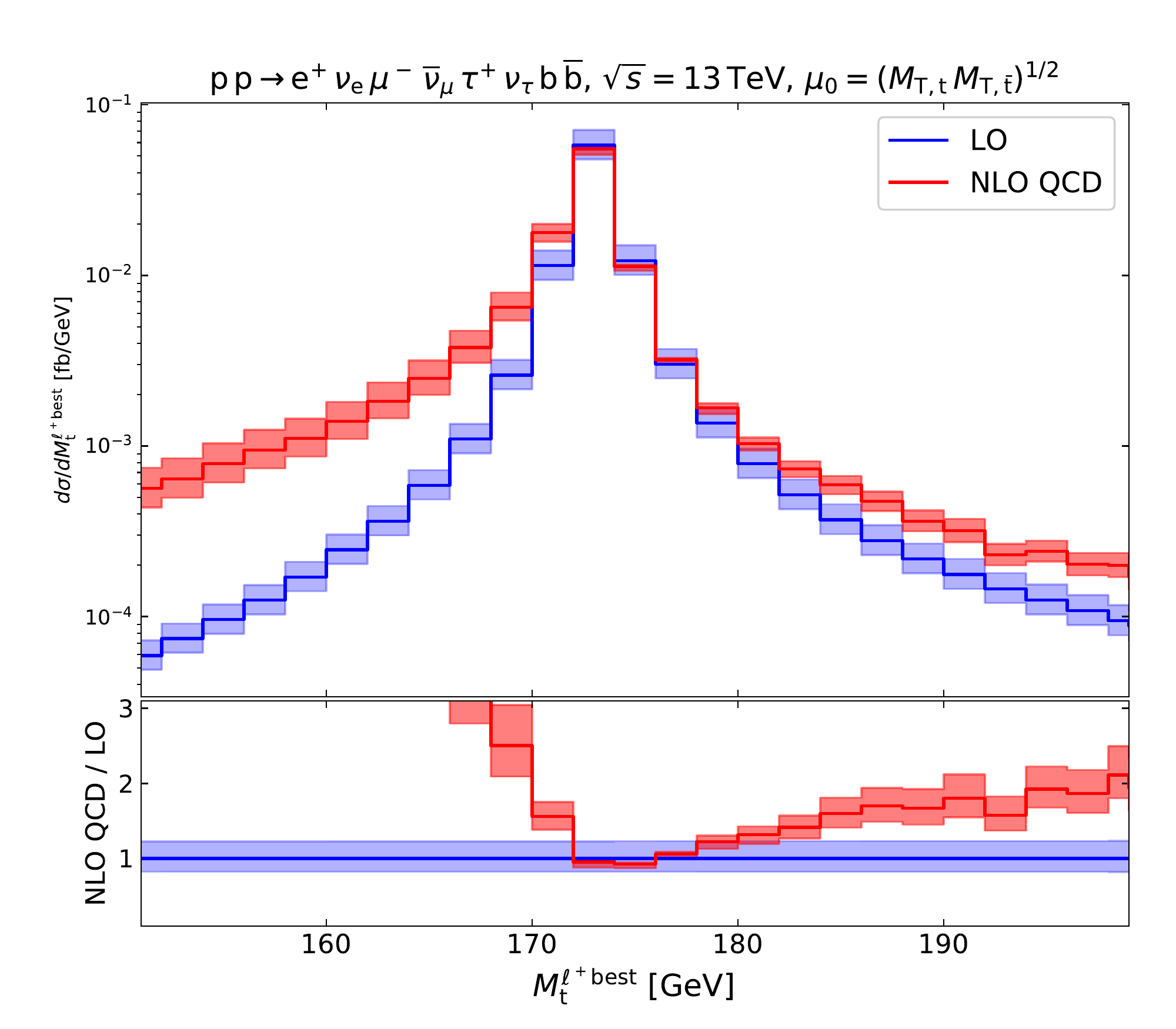}}
  \caption{Same as in \reffi{dphi}.}\label{masstop}
\end{figure}
In \reffis{invantitop_dyn} and \ref{invtop_dyn}, we analyse the
distributions in the invariant masses of the antitop quark and the
reconstructed top quark, respectively. The top quark is reconstructed
according to the above-described requirement that the selected
$\Pl^+\nu_\Pl$ pair gives the invariant mass  closest to $\Mt$ when
combined with the b-quark jet.  In both cases, the NLO corrections are
very large and positive below the top  mass. For a reconstructed
mass of $\Mt-25\GeV$, the NLO cross-section is one order of magnitude
larger than the LO one. This radiative tail is due to final-state
gluon radiation which is not recombined with the decay products of the
antitop or top resonance.  At the peak, the NLO corrections are small
and negative ($-10\%$).  In the off-shell region above the pole mass
they become again large and positive, reaching $\approx+100\%$ at
$\Mt+25\GeV$ for the reconstructed top quark. The relative corrections
are more moderate for the antitop mass ($\approx 70\%$) in the same
region owing to the absence of the reconstruction ambiguity which
affects the top quark in the considered process. The general
features of the correction to this distribution are similar as for
top--antitop production \cite{Denner:2012yc}.

\begin{figure}
  \centering
  \subfigure[Transverse momentum of the reconstructed ${\Pt\overline{\Pt}}$ system.\label{pttt}]{\includegraphics[scale=0.36]{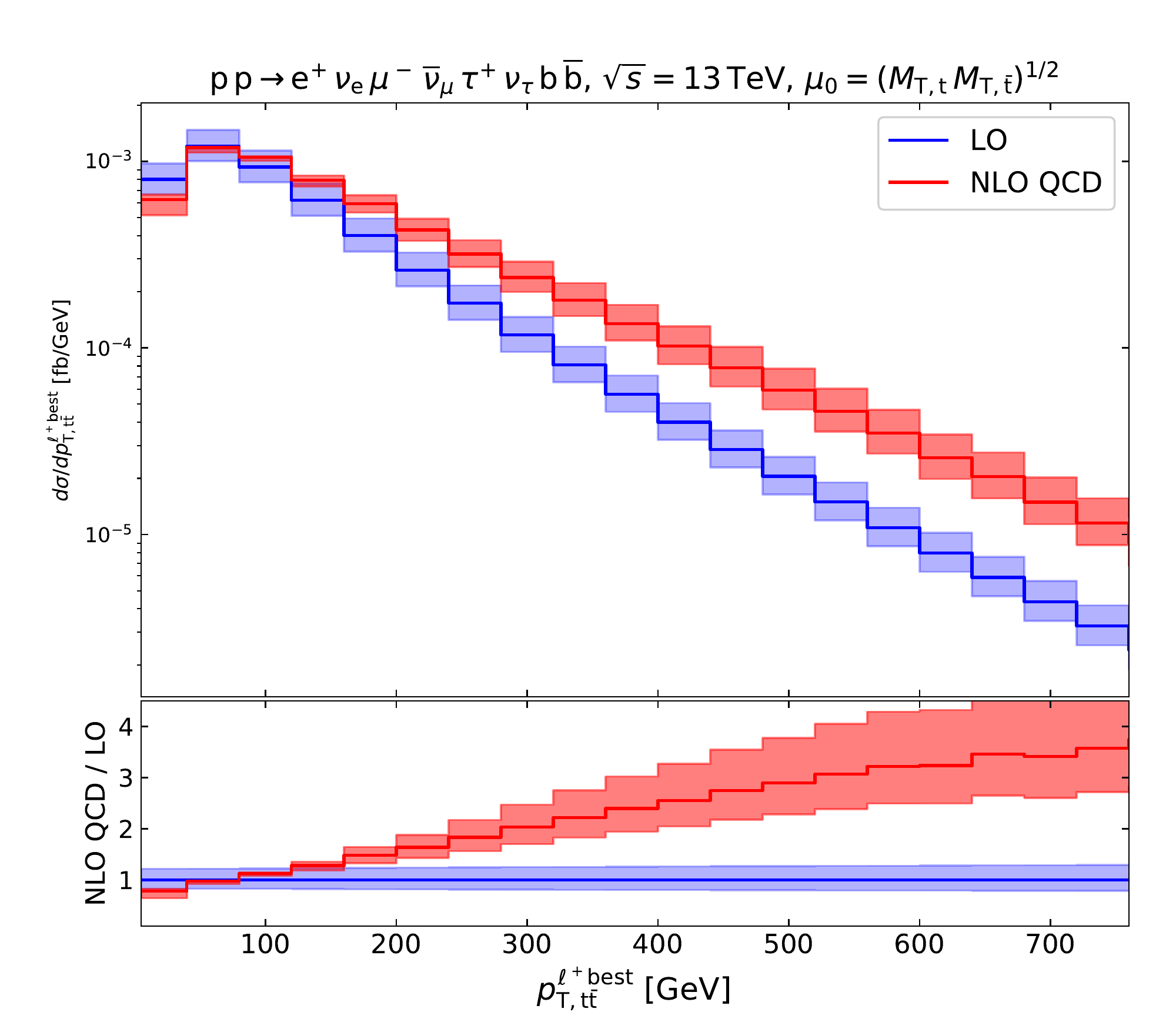}}
  \subfigure[Transverse momentum of the $\Pb\bar{\Pb}$ system.\label{ptbb}]{\includegraphics[scale=0.36]{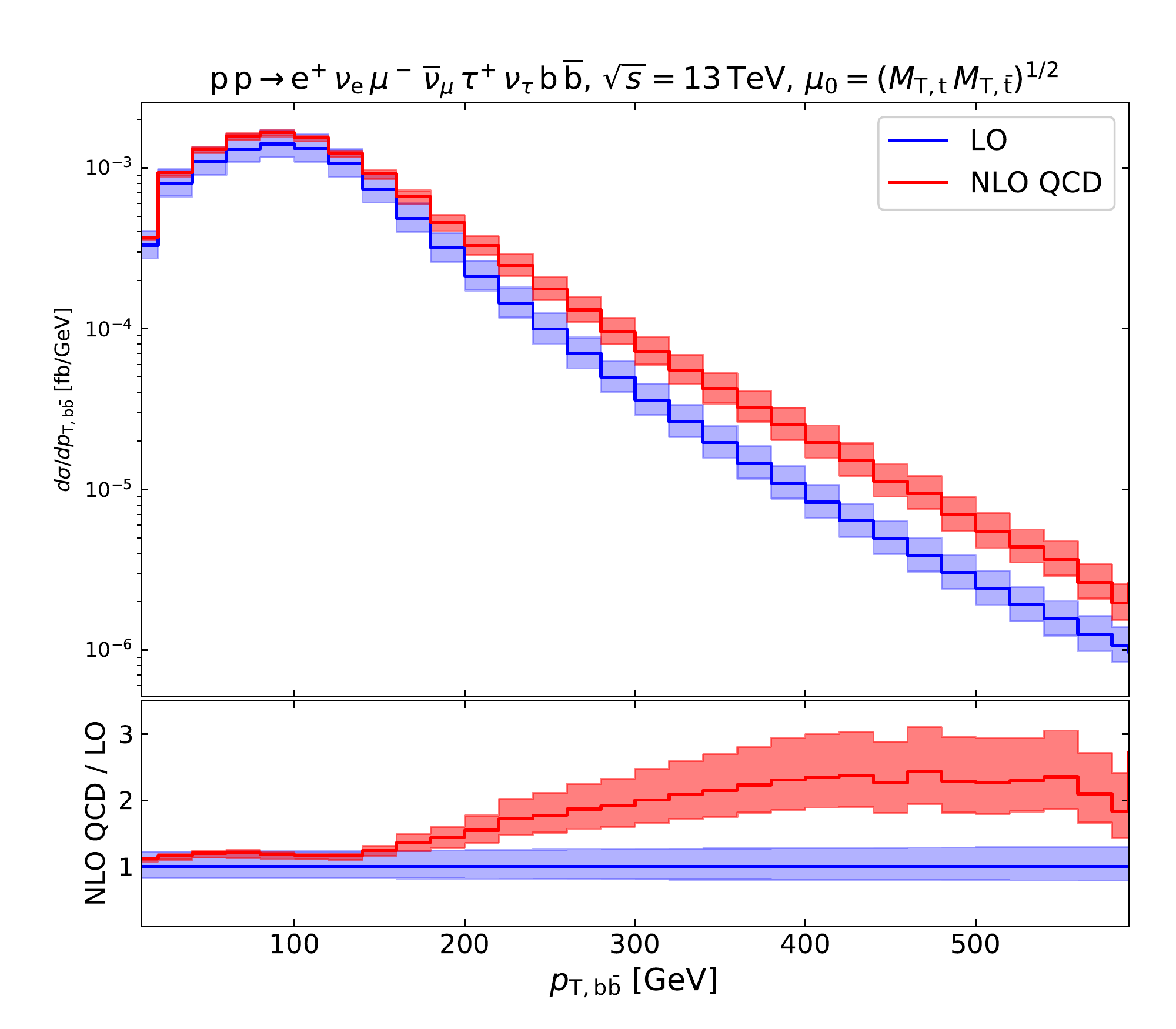}}  
  \caption{Same as in \reffi{dphi}.}\label{othervarbb}
\end{figure}
As already seen in \reffi{masstop}, not all distributions receive NLO
$K$-factors that are close to the value of $1.25$ for the fiducial
cross-section. In \reffi{othervarbb} we consider two more observables
that experience very large NLO corrections in the tails of
distributions, the transverse momentum of the $\Pt\overline{\Pt}$
system (depending on the reconstructed top quark) and the transverse
momentum of the two $\Pb$-jet system shown in \reffi{pttt} and
\reffi{ptbb}, respectively.  The transverse momentum of the
top--antitop system is equivalent to the transverse momentum of the
recoiling system formed by a $\PW^+$ boson and the additional light
jet that may arise at NLO.  The presence of a large real-radiation
contribution explains the huge QCD corrections, which are of order
$+200\%$ for values larger than $500\GeV$. In this region the relative
scale uncertainties are larger than the LO ones, since the dominant
scale uncertainty arises from the real corrections of order
$\alphas^3$ which have only LO accuracy.  In the soft region of the
spectrum the NLO corrections become negative.  The transverse momentum
of the $\Pb\bar{\Pb}$ system (which is measurable at the LHC) is less
directly affected by the additional QCD radiation at NLO. For
$\pt{\Pb\bar{\Pb}}<150\GeV$ the $K$-factor stays almost constant around
$+20\%$, while for larger values it increases, exceeding $+100\%$ for
$\pt{\Pb\bar{\Pb}}>300\GeV$.  Also this large effect is due to the
sizeable contribution of the real radiation which is not clustered
into any of the two $\Pb$~jets. As for the ${\Pt\overline{\Pt}}$
system, the scale uncertainties are large in the tails of the
distribution, as they are dominated by only LO accurate real
radiation. We have checked that the differential $K$-factors and scale 
uncertainties for the two transverse-momentum distributions
practically do not change if either $\mu_0^{\rm (c)}$ or $\mu_0^{\rm
  (e)}$ is used as a central scale.

\subsection{Comparison with double-pole approximation}\label{se:fulvsdpa}
In the previous section we have investigated differential variables
related to the top and antitop resonances of the process as well as
variables that are measurable at the LHC, without having any direct
connection to the dominant resonances.

Since performing the full computation is demanding from the
computational point of view, approximated calculations are often used.
For them, a proper validation is mandatory as they usually rely on
on-shell approximations, which could in principle be far from
reproducing the off-shell structure of the simulated processes.

In this section we compare integrated and differential results
obtained with the full off-shell computation and those obtained with
two different variants of the DPA described in \refse{dpa}. In one
approach, we have applied the DPA to the virtual contributions only,
obtaining impressive agreement with the full results. Alternatively,
we have applied it both to the virtual corrections and to the
$\alphadip$-independent contributions of $I$-operators (see \refse{dpa}).

In Table~\ref{tabledpa} we show the different contributions to the
fiducial NLO cross-sections, computed with full matrix elements and with
the DPA.
\begin{table}
\begin{center}
\begin{tabular}{C{5.5cm}|C{2.6cm}C{2.6cm}C{2.2cm}}
 \cellcolor{blue!9}contribution & \cellcolor{blue!9}full &\cellcolor{blue!9} DPA & \cellcolor{blue!9}$\Delta_{\rm DPA}$\\
 \hline\\[-0.35cm]
 {B} (Born, with NLO PDFs) & $~0.20326(4)$ & $~0.19770(3)$ & $-2.7\%$ \\[0.1cm]
 {V} (virtual) & $ -0.02079(5)$ & $-0.02071(3)$ & $-0.4\%$ \\[0.1cm]
 {I} (integrated dipoles, \\ ${\alphadip}$-independent $I$-operator) & $-0.1004(1)$ & $-0.0976(1)$ & $-2.8\%$\\[0.1cm]
 {I} (integrated dipoles, \\ ${\alphadip}$-dependent $I$-operator, ${\alphadip}=0.01$) & $-0.3276(3)$ & $-0.3204(3)$ & $-2.2\%$\\[0.1cm]
% {I} (integrated dipoles, \\ ${\alphadip}$-dependent $I$-op. + $K,P$-op.) & $-0.1594(2)$ & -- & -- \\[0.1cm]
 {I} (integrated dipoles, $K,P$-operators, ${\alphadip}=0.01$) & $~0.1682(2)$ & -- & -- \\[0.1cm]
% {I} (integrated dipoles, complete $I$-op., $\alphadip=0.01$) & $-0.4280(4)$ & $-0.4178(3)$ & $-2.4\%$ \\[0.1cm]
 {R} (real subtracted) & $~0.3167(5)$ &-- & -- \\[0.1cm]
 \hline
\end{tabular}
\end{center}
\caption{Contributions to the fiducial NLO cross-sections (fb), obtained
  with the off-shell calculation and with the DPA.
The central scale $\mu_0^{\rm (d)}$ is understood. Numerical
uncertainties are shown in parentheses. The value $\Delta_{\rm DPA}$ is
computed as the difference between the DPA and the full results normalised to the full one.}
\label{tabledpa}
\end{table}
The DPA underestimates the Born cross-section by $2.7\%$.  Since the
deviations from the full results can be very large in some kinematic
regions, we are not showing differential distributions at Born level.
The virtual contributions are very well described by the DPA, while
the $\alphadip$-independent and $\alphadip$-dependent terms of the
$I$-operators (which take negative values) are reproduced with $2.8\%$
and $2.2\%$ negative deviation, respectively, similarly to the LO
contributions and of the expected order $\Gt/\Mt$.

The fiducial NLO cross-sections obtained with the full off-shell and the
two DPA calculations read:
\beq
\sigma_{\rm NLO}^{\rm full} = 0.2394(6)^{+5.4\%}_{-7.2\%} \fb\,,\quad \sigma_{\rm NLO}^{\rm DPA,\, V} = 0.2395(7)^{+5.8\%}_{-7.4\%} \fb\,,\quad \sigma_{\rm NLO}^{\rm DPA,\, V+I} = 0.2422(7)^{+6.3\%}_{-7.6\%}\,. \nnb
\eeq
The uncertainty of the DPA approximation can be estimated by
multiplying the generic relative uncertainty of the DPA of $\mc
O(\Gt/\Mt)$ with the absolute size of the terms that are
approximated. Since the finite virtual corrections contribute less
than $10\%$ to the NLO fiducial cross-section, the DPA uncertainty of
$\sigma_{\rm NLO}^{\rm DPA,\, V}$ is of the order $10\%\times \Gt/\Mt
\approx 0.1\%$. The difference to the full results turns out to be
somewhat smaller.  Since the integrated dipoles make up $50\%$ of the
NLO cross-section, treating them in DPA yields an uncertainty of the
order of $50\%\times \Gt/\Mt \approx 0.5\%$. The cross section
$\sigma_{\rm NLO}^{\rm DPA,\, V+I}$ differs from the full NLO one by
$1.2\%$, which is of the same order of magnitude.
Treating also the $\alphadip$-dependent integrated-dipole
contributions within the DPA, would deteriorate the accuracy even more
since these contributions are even larger but cancel to a large extent
with the real subtracted contributions that are not treated with the
DPA (see \refta{tabledpa}). 
This option is not considered any further in this paper.
The percent-level discrepancy between $\sigma_{\rm NLO}^{\rm DPA,\,
  V+I}$ and $\sigma_{\rm NLO}^{\rm full}$ translates to even larger
discrepancies when looking at exclusive phase-space regions.
Note that in both DPA calculations, the scale variations
give slightly larger scale uncertainties than in the full
calculation.

\begin{figure}[t]
  \centering
  \subfigure[Transverse momentum of the $\Pb\bar{\Pb}$ system. \label{dpa:ptbb}]{\includegraphics[scale=0.36]{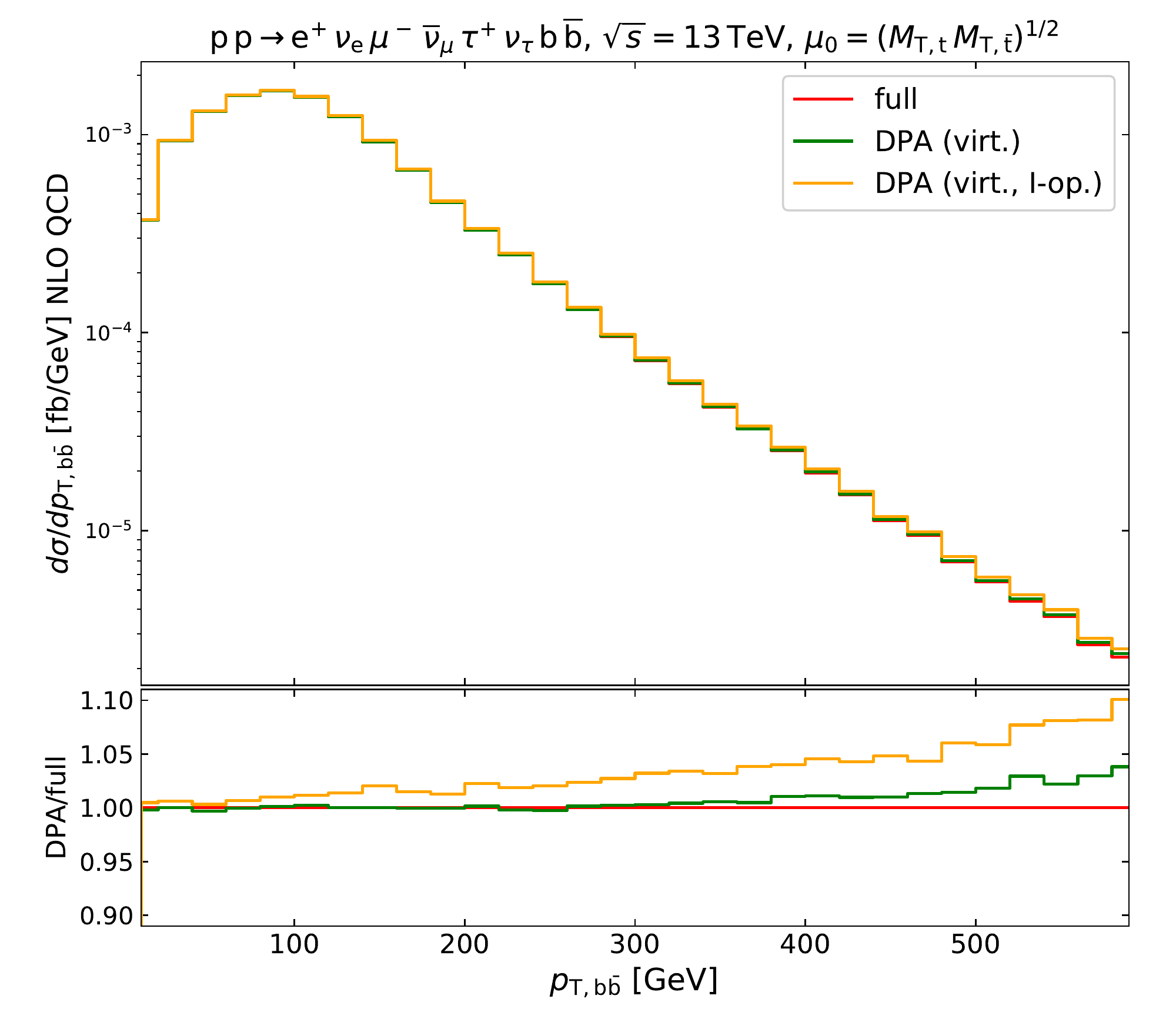}}
  \subfigure[Invariant mass of the $\Pb\bar{\Pb}$ system. \label{dpa:mbb}]{\includegraphics[scale=0.36]{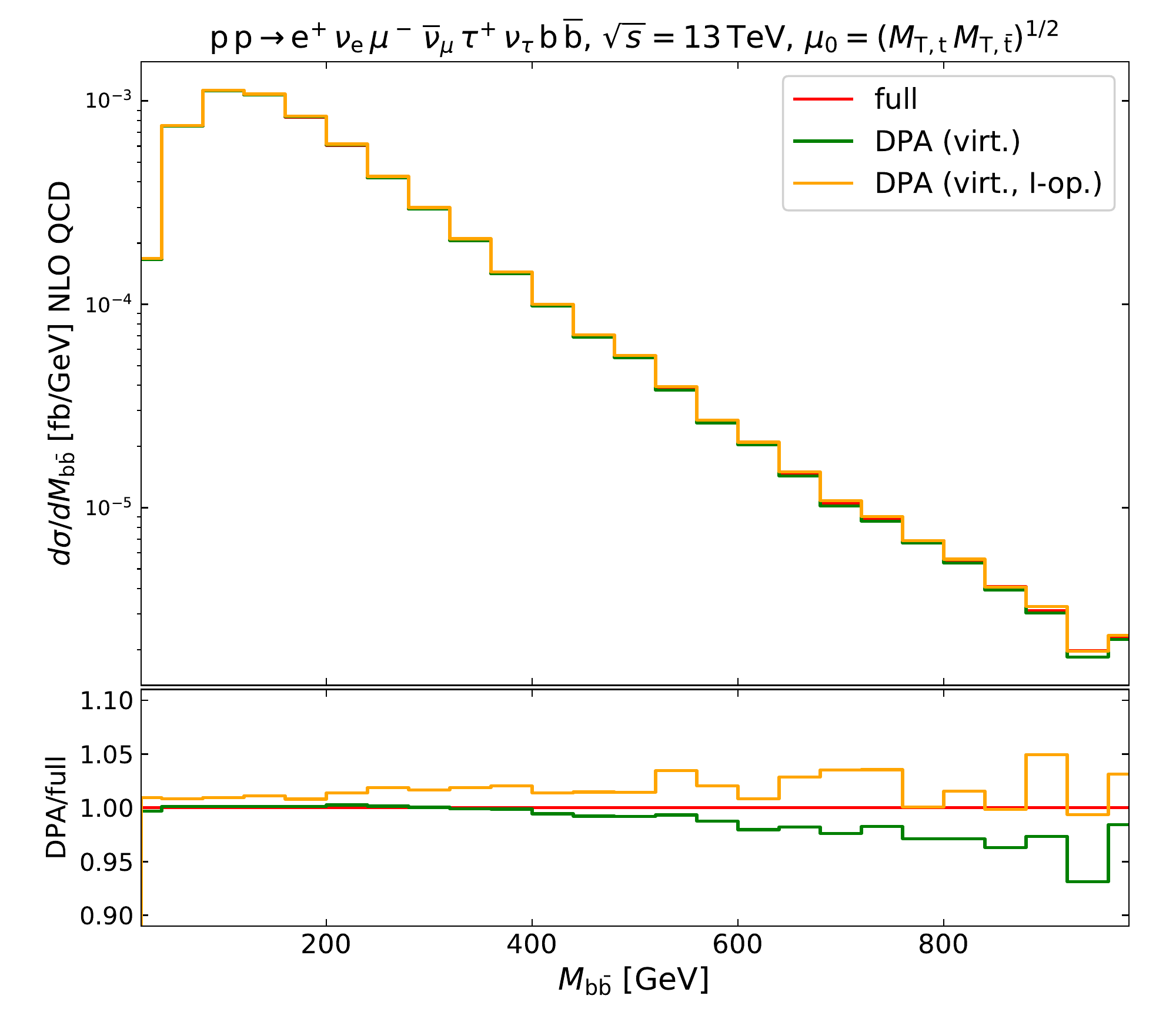}}
  \caption{Differential distributions at NLO QCD, 
    obtained with full matrix elements, and applying the DPA either to
    the virtual contributions only, or both to the virtual and to the
    $\alphadip$-independent part of $I$-operators.  The dynamical scale
    $\mu_0^{\rm (d)}={\left(\tm{\Pt}\,\tm{\bar{\Pt}}\right)}^{1/2}$ is
    understood.  Differential cross-sections are shown in the upper
    plot, the ratio over the full result in the lower
    plot.}\label{dpa:bb}
\end{figure}

In \reffi{dpa:bb} we consider the transverse-momentum and 
invariant-mass distributions of the two-$\Pb$-jet system. 
Discrepancies at the level of $5\%$ can be found in the tails of the
two distributions, though with opposite signs, when approximating the
virtual corrections. The DPA gives a larger NLO cross-section in the
large-transverse-momentum regime and a smaller one for large invariant
masses.  In \citere{Bevilacqua:2020pzy}, an analogous effect of even
$20\%$ has been found in the invariant-mass spectrum using the
narrow-width approximation.  Applying the DPA also to the integrated
dipoles shifts the results towards larger values in the tails of both
distributions, resulting in a $+10\%$ deviation for
$\pt{\Pb\bar{\Pb}}=600\GeV$, and in a deviation of $2\%$ at large
invariant mass ($M_{\Pb\bar{\Pb}}>200\GeV$).  In the tails of the
distributions, the relative contribution of integrated dipoles to the
cross-section is even larger than at the integrated level.

\begin{figure}[t]
  \centering
  \subfigure[Invariant mass of the positron--electron-neu\-tri\-no--bottom
  system. \label{dpa:mtope}]
{\includegraphics[scale=0.36]{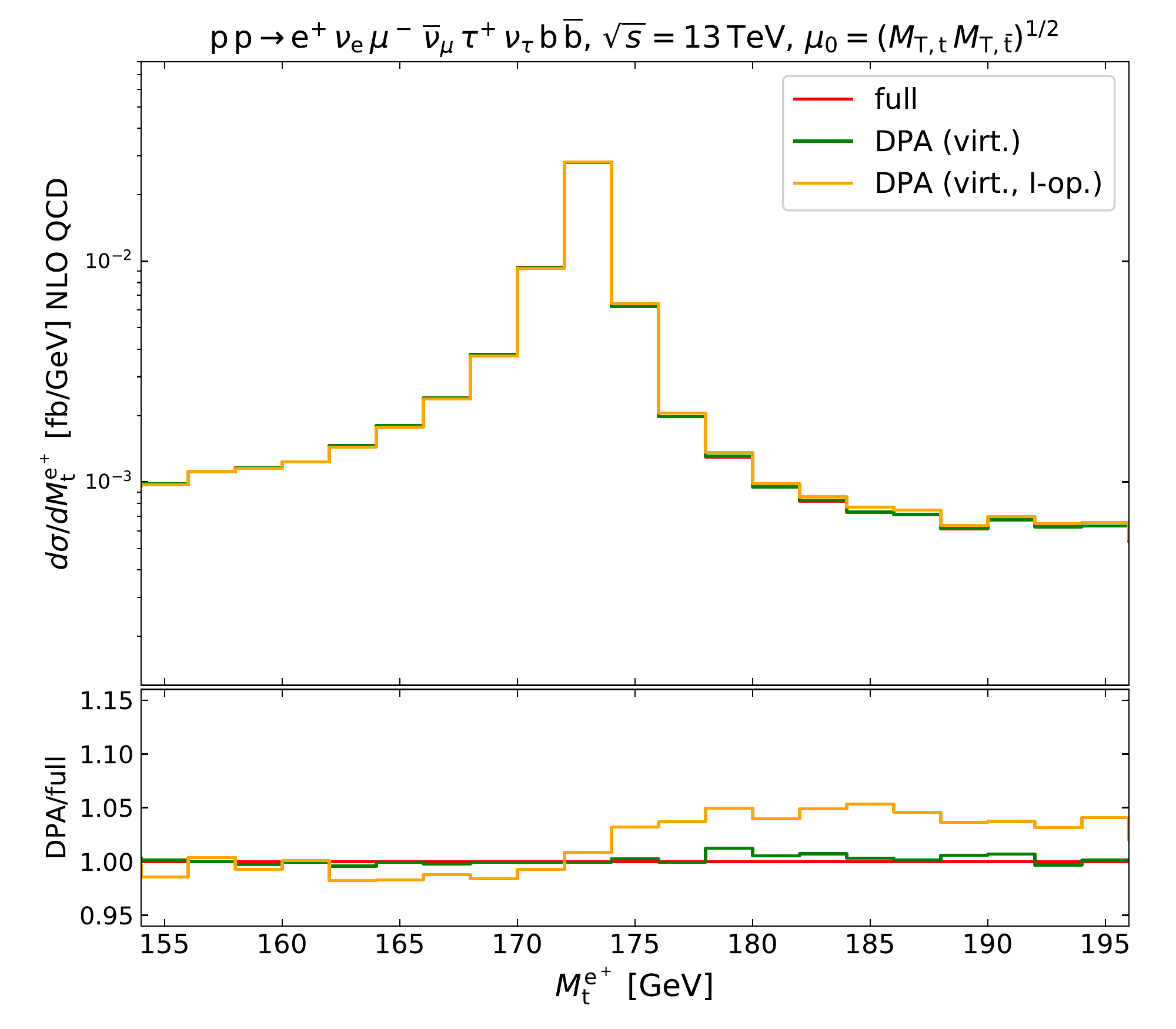}}
  \subfigure[Invariant mass of the reconstructed top quark.\label{dpa:mtopl}]{\includegraphics[scale=0.36]{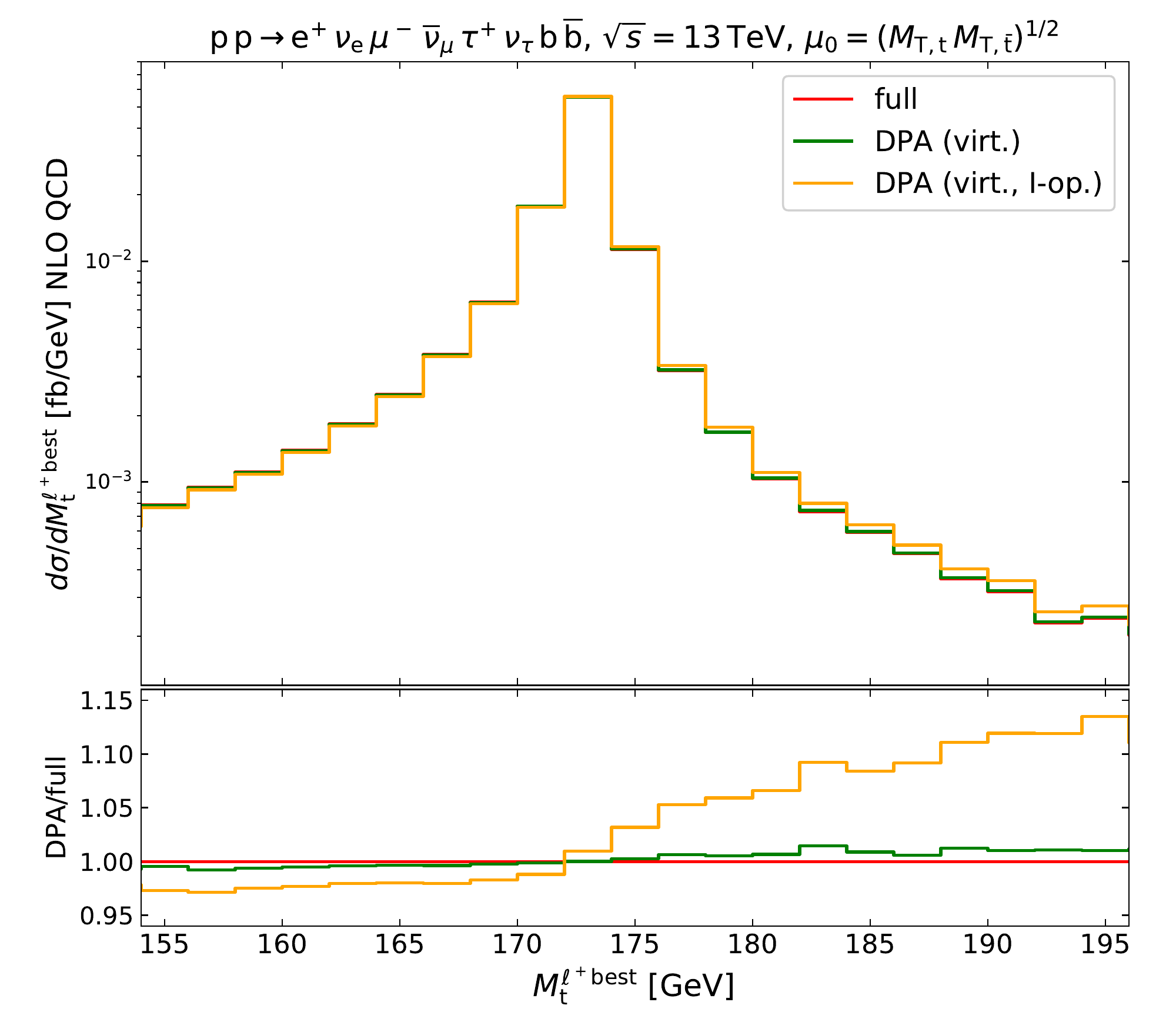}}
  \caption{Same as in \reffi{dpa:bb}.}\label{dpa:mtop}
\end{figure}

To investigate how well the top-quark resonance is reproduced within
the DPA, we consider the distribution in the invariant mass of the
positron--electron-neutrino--bottom system, $M_{\Pe^+\nu_\Pe\Pb}$, and
of the reconstructed (with the same criterion as used in the
previous sections) top quark in \reffi{dpa:mtop}.
The DPA applied to virtual corrections only leads to a result that is
slightly smaller (larger) than the full one for invariant masses below
(above) the top-quark mass. However, this effect is very small (at
most $1\%$) for both variants of the reconstructed top quark mass.
Applying the DPA to integrated dipoles affects the two distributions
differently for values larger than $\Mt$.  For $M_{\Pe^+\nu_\Pe\Pb}$,
the DPA result is $5\%$ larger than the full result in this region,
while choosing the reconstructed top-quark mass gives a larger
deviation reaching $+13\%$ for $M_{\Pt}^{\Pl^+\,\rm best}=\Mt +
20\GeV$. The large deviations appear in the off-shell region above the
top resonance in which the resonant contributions are suppressed.
The effect is smaller in \reffi{dpa:mtope} because of the background
of about half of the events, where the resonant top-quark decays to
$\Pb\tau^+\nu_\tau$. 

\begin{figure}[t]
  \centering
  \subfigure[$H_{\rT}$ variable, excluding additional radiation.\label{dpa:ht}]{\includegraphics[scale=0.36]{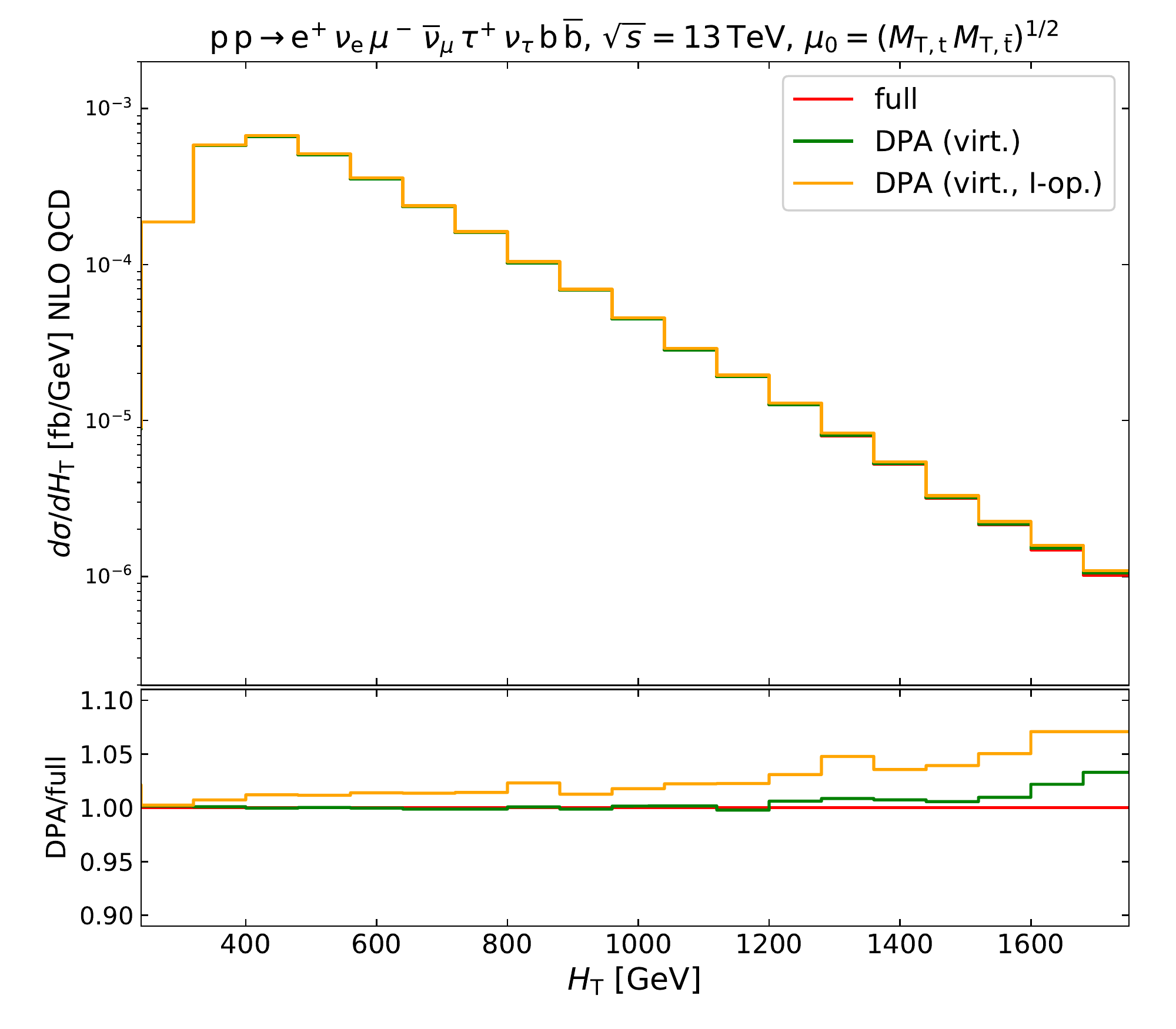}}
  \subfigure[Transverse momentum of the antitop quark.\label{dpa:ptt}]{\includegraphics[scale=0.36]{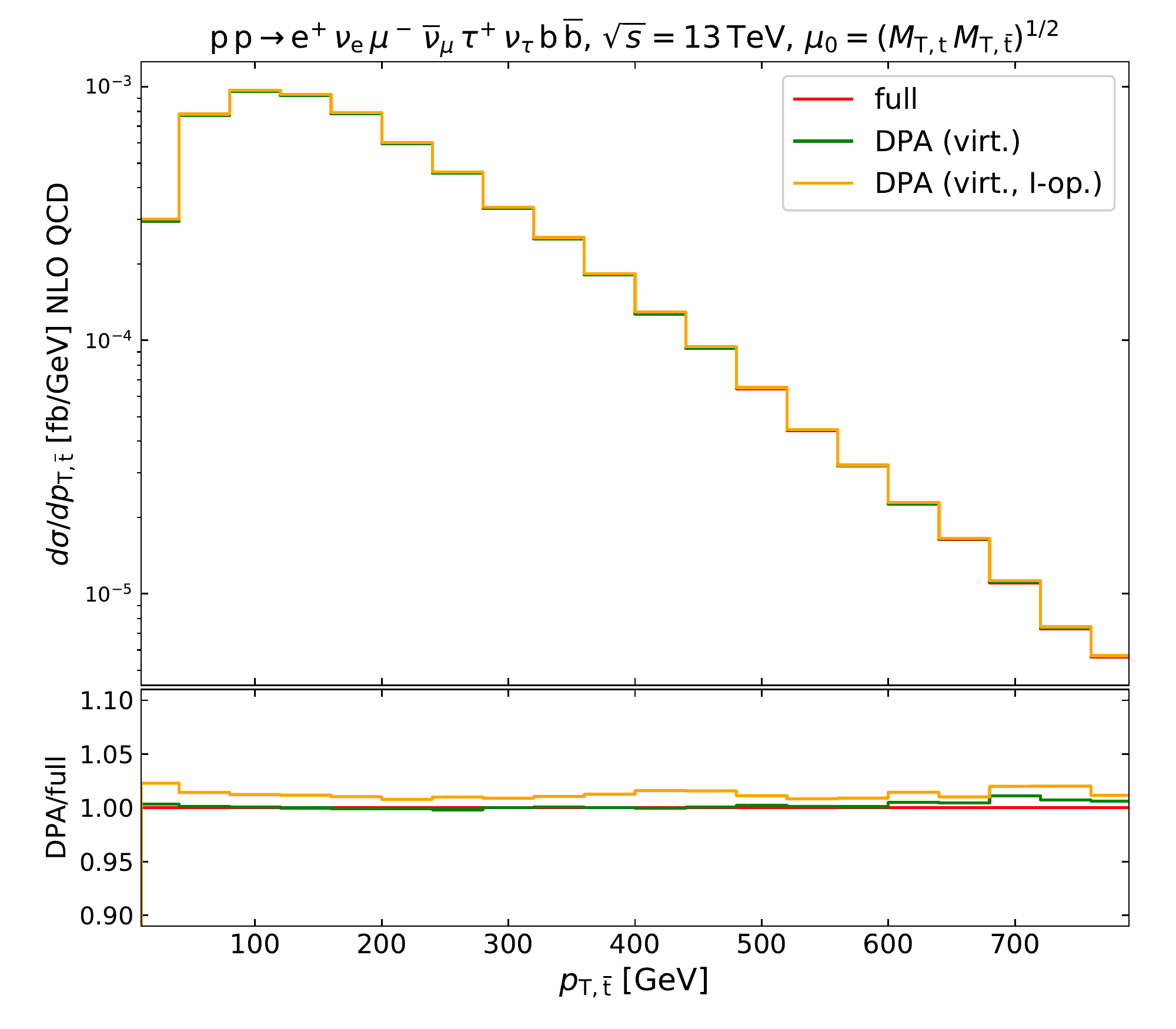}}  
  \caption{Same as in \reffi{dpa:bb}.}\label{dpa:other}
\end{figure}

The distribution in the $H_{\rT}$ observable is considered in
\reffi{dpa:ht}.
The DPA calculation gives a relevant enhancement to the off-shell
result in the tail of the distribution. The size of this effect for
$H_{\rT}> 1600\GeV$ is $3\%$ if the DPA is applied only to the virtual
contributions and $7\%$ if it is applied also to the integrated dipoles.
The deviation has some similarities with the one observed in the
$\Pb\bar{\Pb}$ system transverse momentum [see \reffi{dpa:ptbb}].  The
$H_{\rT}$ variable has been studied in \citere{Bevilacqua:2020pzy},
finding that the narrow-width approximation underestimates the full
cross-section for large $H_{\rT}$ by some $30\%$, going in the opposite direction
with respect to the DPA technique we consider here.  This highlights
that approximated computations should be taken with a grain of salt,
and must be verified with results based on full matrix elements, in
particular,  in regions of phase space where non-doubly-resonant
contributions are not suppressed.

As a last example, we show the distributions in the antitop-quark
transverse momentum in \reffi{dpa:ptt}.  This variable is well
described by the DPA in the whole spectrum, even in the tail.  The
discrepancy between the two DPA curves and the full one is roughly
constant and reflects the results at the integrated level.  This
illustrates that the DPA works very well also at the differential
level for kinematic variables that are directly related to the
antitop or top quark.  
Both for the DPA applied to only the virtual corrections and
the DPA applied also to the contributions of the $I$-operators the
relative deviations from the off-shell results
are flat and of the same size as for the fiducial
cross-section.
%from the full calculation amount to $1$--$2\%$, as expected
%from the intrinsic uncertainty of the DPA [$\cO (\Gt/\Mt)$].

\section{Conclusions}\label{sec:conclusions}

In this paper, we have presented the calculation of NLO QCD
corrections to the off-shell production of a top--antitop-quark pair
in association with a $\PW^+$ boson, considering a final state with
three charged leptons, two $\Pb$~jets and missing energy. Although our 
results have been obtained for the specific case of three different
lepton flavours, they hold to good approximation also for the case of
two identical positively-charged leptons up to a global symmetry
factor 1/2.  All non-resonant effects, interferences, as well as spin
correlations are accounted for in all parts of the computation. We
have studied both integrated and differential cross-sections, and
performed a comparison between the results obtained from full matrix
elements and those obtained using a double-pole approximation (DPA).

We have investigated different choices of the central factorisation
and renormalisation scale, both fixed and dynamical. We have found
that employing a dynamical scale gives better-behaved NLO QCD predictions
than using the fixed scale $\mu_0=\Mt+\Mw/2$. This holds both for
a scale based on the transverse-momentum content of the final state,
$\mu_0=H_{\rT}$, and for a scale based on the transverse masses
of the dominant top--antitop resonances,
$\mu_0=(M_{\Pt,\rT}\,M_{\bar{\Pt},\rT})^{1/2}$. 

Considering proton--proton collisions at a $13\TeV$, the QCD corrections
with a dynamical scale choice are positive and moderate ($+25\%$) at
the integrated level. The theoretical uncertainties estimated from
scale variations decrease from $20\%$ at LO to $5\%$ at NLO.
Corrections for distribution become larger than those for the fiducial
cross-section in several phase-space regions, in particular, in the
tails of energy-dependent distributions which depend on the hadronic
activity ($\Pb$~jets, additional QCD radiation), both for measurable
LHC observables and for variables based on Monte Carlo truth
describing the top and antitop resonances. The scale uncertainties
are of the same or even larger size than the LO ones in the tails of
certain transverse-momentum and invariant-mass distributions.

The DPA reproduces in a surprisingly accurate manner most results of
the calculation based on full matrix elements if applied only to the
virtual corrections, thanks to the relatively small size of such
contributions.  When the DPA is applied also to the integrated
subtraction counterterms ($I$-operators in the Catani--Seymour
scheme), it reproduces the full results at the $1\%$ level for the
fiducial cross-section and top-dependent observables, while deviations
of a few percent are found in more exclusive parts of the phase space,
in agreement with the intrinsic uncertainty of the approximation.
However, in regions of phase space that are not dominated by
top--antitop production, the discrepancy can reach $10\%$ and more.

In conclusion, the calculation presented in this paper represents an
important step towards a precise theoretical description of
${\Pt\overline{\Pt}\PW}$ production, which will be beneficial both in
direct searches and in the background modelling for other relevant
signals at the LHC.  Since off-shell $\Pt\overline{\Pt}\PW$ production
is dominated by contributions from top and antitop resonances, the approximated
calculation relying on the resonant contributions is expected to give
a reasonable picture of the process in sufficiently inclusive
phase-space regions.  However, given the increased luminosity that
will be available in the next LHC runs, in view of a precise and fully
trustworthy comparison between experimental data and theoretical
predictions, the calculation based on full matrix elements is strongly
recommended, as it gives a more complete description of
the process in inclusive and exclusive kinematic regions.

\section*{Acknowledgements}
We are grateful to Timo Schmidt and Mathieu Pellen for help with
\mocanlo and to Jean-Nicolas Lang and Sandro Uccirati for maintaining
\recola.  This work is supported by the German Federal Ministry for
Education and Research (BMBF) under contract no.~05H18WWCA1.

\bibliographystyle{JHEP}
\bibliography{ttv}

%%%%%%%%%%%%%%%%%%%%%%%%%%%%%%%%%

\end{document}